\documentclass{pasj00}
\begin{document}

\author{Mariko \textsc{Kimura}, \altaffilmark{1}}
\email{mkimura@kusastro.kyoto-u.ac.jp}

\author{Shin \textsc{Mineshige}, \altaffilmark{1}}
\email{shm@kusastro.kyoto-u.ac.jp}
\and

\author{Norita \textsc{Kawanaka}, \altaffilmark{2}}
\email{norita@astron.s.u-tokyo.ac.jp}

\altaffiltext{1}{Department of Astronomy, Graduate School of Science, Kyoto University, Oiwakecho, Kitashirakawa, Sakyo-ku, Kyoto 606-8502}
\altaffiltext{2}{Department of Astronomy, Graduate School of Science, The University of Tokyo, 7-3-1, Hongo, Bunkyo-ku, Tokyo 113-0033}

\title{How does a Secular Instability Grow in a Hyperaccretion Flow?}

\Received{} \Accepted{}


\SetRunningHead{Kimura et al.}{Time Evolution of a unstable NDAF}

\maketitle

\begin{abstract}
Hyperaccretion flows with mass accretion rates far above the Eddington rate have an N-shaped equilibrium curve on the $\Sigma$-$\dot{M}$ plane (with $\Sigma$ and $\dot{M}$ being surface density and mass accretion rate, respectively).
The accretion flow on the lower $\Sigma$ branch of the N-shape is optically thick, advection-dominated accretion flow (ADAF) while that on the upper one is neutrino-dominated accretion flow (NDAF).
The middle branch has a negative slope on the $\Sigma$-$\dot{M}$ plane, meaning that the flow on this branch is secularly unstable.
To investigate how the instability affects the flow structure and what observable signatures are produced, we study the time evolution of the unstable hyperaccretion flow in response to variable mass injection rates by solving the height-averaged equations for viscous accretion flows.
When a transition occurs from the lower branch to the upper branch (or from the upper branch to the lower branch), the surface density rapidly increases (decreases) around that transition region, which induces locally enhanced mass flow (referred to as non-steady mass flow) into (out of) that region.
We confirm that the non-steady flow can create a kind of disturbance and that it propagates over the whole disk.  However, the non-steady mass flow is not strong enough to induce coherent transition over the whole disk, unless the mass injection rate varies with time.  When the injection rate continuously changes, the neutrino luminosity varies intermittently, thus producing step-function-like light curves, as the radiation efficiency discontinuously changes every time the local transition occurs.
The effects of changing the N-shape and possible observational consequences on the gamma-ray burst emission are briefly discussed in relation to gamma-ray bursts.
\end{abstract}

\section{Introduction}

Gamma-ray bursts (GRBs) are the brightest explosions in the universe. They release energy up to $\gtrsim 10^{51}~{\rm ergs}$, which is on the order of $10^{-3}M_{\solar}c^{2}$, in the form of gamma-rays within a few seconds.  From the spectral features and light curves of their prompt and afterglow emission, they are considered to be produced in ultrarelativistic jets.  Although the picture of the central engine has not yet been established, several constraints are required from the observations.  It should be able to launch ultrarelativistic jets, meaning that relativistic objects should be involved.  It should have two different characteristic timescales: the variability timescale of their prompt emission ($\sim 10^{-3}-10^{-1}~{\rm s}$, e.g., \cite{beloborodov+00, ackermann+10, troja+15}) and the duration timescale of the prompt emission ($\sim 2 - 1000~{\rm s}$).  The latter timescale is much longer than the dynamical timescale of relativistic objects, but there is a way to explain all the requirements.  The most promising model of the GRB central engine is a hyperaccretion flow \citep{narayan+92, narayan+01}.  In this model an ultrarelativistic jet would be launched from a massive accretion disk around a stellar-mass black hole whose mass accretion rate is at most $\sim 0.01 - 1M_{\odot}~{\rm s}^{-1}$.  Such a system is expected to form after a cataclysmic event such as the gravitational collapse of a massive star or the merger of a neutron star - neutron star binary or a black hole - neutron star binary \citep{piran99}.  

Since such an accretion flow is extremely optically thick with respect to photons, it cannot cool via electromagnetic radiation.  Instead, since the density and temperature of this accretion flow are very high ($\rho \gtrsim 10^7~{\rm g}~{\rm cm}^{-3}$, $T \gtrsim 10^{10}~{\rm K}$), it would cool via neutrinos, and such an accretion flow is often called as ``neutrino-dominated accretion flows'' (NDAF; \cite{popham+99, narayan+01, dimatteo+02, kohrimineshige02, kohri+05, gu+06, chenbeloborodov07, kawanakamineshige07, liu+07, kawanakakohri12, xue+13}; see also section 10.6 of \cite{kato+08}).  In this model, the variability timescale can be interpreted as the dynamical (i.e., rotation) timescale of an accretion disk, while the burst duration timescale corresponds to the accretion timescale.  As for how a hyperaccretion flow can produce an ultrarelativistic jet, two physical processes are mainly discussed in the literature: neutrino pair annihilation ($\nu\bar{\nu}\rightarrow e^-e^+$; \cite{eichler+89, asanofukuyama00, zalameabeloborodov11,suwa13}) and magnetohydrodynamic (MHD) mechanisms such as the Blandford-Znajek (BZ) process (\cite{blandfordznajek77, mckinneygammie04, hawleykrolik06}).

In either scenario, it is implied that the energy deposition rate (corresponding to the jet luminosity) as a function of mass accretion rate has a step-function-like behavior at a certain critical mass accretion rate.  According to the neutrino pair annihilation scenario, in fact, we expect a big jump in the neutrino emission efficiency across the critical accretion rate, $\dot{M}_{\rm ign}\sim 0.003 - 0.01M_{\odot}~{\rm s}^{-1}$ for fiducial parameters (see \cite{kawanaka+13a}).  That is, when the mass accretion rate is above $\dot{M}_{\rm ign}$ the accretion flow becomes an efficient neutrino emitter (``NDAF'' regime) and the energy deposition via neutrino annihilation is efficient, while when $\dot{M}$ is below $\dot{M}_{\rm ign}$ the accretion flow is in the ``advection-dominated'' (ADAF) regime and neutrino pair annihilation emission is not at all efficient.  Similar behavior is expected in the MHD scenario, as well.  \citet{kawanaka+13a} investigated the MHD jet luminosity predicted from a hyperaccretion flow and found that it shows step-function-like behavior at $\dot{M}\sim \dot{M}_{\rm ign}$.  In either case, therefore, one can expect a drastic change of jet luminosity from a hyperaccretion flow when its mass accretion rate attains the critical value $\sim \dot{M}_{\rm ign}$.  

If there is any physical mechanism which can modulate the mass accretion rate around $\dot{M}_{\rm ign}$, the jet luminosity would change violently, which may explain to the highly variable GRB emission.  Theoretically, several types of disk instabilities are known \citep{lightmaneardley74, shibazakihoshi75, shakurasunyaev76, piran78}, and the large-amplitude luminosity variations of accreting systems, such as those observed in dwarf novae and X-ray binaries are often ascribed to the accretion disk instabilities (e.g., \cite{osaki74, mineshigewheeler89, honma+91}, see a concise review by \cite{kato+08} chapter 5 and 10).
However, little was known about the disk instabilities of a hyperaccretion flow, especially that with mass accretion rate around $\dot{M}_{\rm ign}$.  Recently, \citet{kawanaka+13b} found that the thermal equilibrium curve at the innermost region of a hyperaccretion flow on the $\Sigma$-$\dot{M}$ plane has the structure like a character ``N'', which has a negative slope at $\dot{M}\sim \dot{M}_{\rm ign}$.  This means the existence of a secular (or viscous) instability of the accretion flow.  

In our study, we for the first time solve the time evolution of the unstable hyperaccretion flow and show that the luminosity evolution occurs intermittently as a consequence of a local transition which takes place in a narrow region.  The plan of this paper is as follows: in Section 2, we describe the equations for disk evolution and the modeled N-shaped thermal equilibrium curve.  In Section 3 we show the results of our calculations with an emphasis on the consequence of the instability.  Section 4 is devoted to the discussion on the dependence on the N-curve, the observable behavior, the comparison with the case of S-shaped curve, and future issues.

\section{Methods of Calculation}

\subsection{Basic Equations for Viscous Diffusion}

In the present study, we solve the height-averaged basic equations for viscous diffusion of an axisymmetric accretion disk (see, e.g., \cite{kato+08} chapter 3):

\begin{equation}
\frac{\partial \Sigma}{\partial t} = \frac{1}{2\pi r} \frac{\partial \dot{M}}{\partial r},
\label{basic1}
\end{equation}
\noindent
and
\begin{equation}
\dot{M} \left[\frac{d}{dr}(r^{2}\Omega) \right] = -2\pi \frac{\partial}{\partial r} \left(r^{3} \nu \Sigma \frac{d\Omega}{dr} \right).
\label{basic2}
\end{equation}
Here, $\Sigma(r) \equiv \int \rho(r, z) dz$ (with $\rho$ being mass density) is the surface density, $r$ is the distance from the center of a black hole, $\dot{M} \equiv -2\pi r \Sigma v_r$ (with $v_{r}$ being the radial velocity), $\nu$ is the kinematic viscosity, and $\Omega$ is the angular velocity.
Equation (\ref{basic1}) describes mass conservation, while equation (\ref{basic2}) represents angular momentum conservation.
In our calculations, we adopt the pseudo-Newtonian force for a Kerr black hole (Artemova et al. 1996) for calculating $\Omega$ in equation (\ref{basic2}); that is, from

\begin{equation}
F = -\frac{GM_{\rm BH}}{r^{2-\beta}(r-r_{\rm H})^{\beta}}.
\label{pseudo}
\end{equation}
We find
\begin{equation}
\Omega = \sqrt{\mathstrut{\frac{GM_{\rm BH}}{r^{3-\beta} (r - r_{\rm H})^{\beta}}}}.
\label{Omega}
\end{equation}
where $M_{\rm BH}$ is the black hole mass, $r_{\rm H}$ is the radius of the event horizon, $\beta \equiv (r_{\rm ms}/r_{\rm H})-1$ (with $r_{\rm ms}$ being the radius of the marginally stable orbit) is a constant to be specified later, and we assumed the balance between the gravitational force and the centrifugal force.
Inserting equation (\ref{Omega}) to equation (\ref{basic2}), we obtain

\begin{equation}
\dot{M} = 6\pi \frac{(r-r_{\rm H})^{\frac{\beta+2}{2}}}{\left[r-(1+\beta)r_{\rm H}\right] r^{\frac{\beta-1}{2}}} \frac{\partial}{\partial r} \left[\frac{r^{\frac{\beta+1}{2}}(r-\frac{3-\beta}{3}r_{\rm H})}{(r-r_{\rm H})^{\frac{\beta+2}{2}}} \nu \Sigma \right].
\label{basic2-pk}
\end{equation}

We next transform equations (\ref{basic1}) and (\ref{basic2-pk}) to non-dimensional forms.
For this purpose we introduce the following dimensionless variables:
\begin{eqnarray}
x \equiv \sqrt{\mathstrut{r/r_{\rm out}}}, \\
x_{\rm H} \equiv \sqrt{\mathstrut{r_{\rm H}/r_{\rm out}}}, \\
\sigma \equiv \Sigma/\Sigma_0, \\
\tau \equiv t/t_0 \quad \mbox{with} \quad t_{0} \equiv r_{\rm out}^{2} \Sigma_{0} / \dot{M}_{0}, \\
\dot{m} \equiv \dot{M}/\dot{M}_0, \\
\mu \equiv \nu\Sigma / \dot{M}_{0}.
\label{mujigen}
\end{eqnarray}
Here, $r_{\rm out}$ is the size of the innermost region of the disk, which we are concerned with, $\Sigma_0$ and $\dot{M}_{0}$ represent typical surface density and mass accretion rate of NDAF, and $t_0$ corresponds to the viscous timescale at the outer edge of the disk. These normalization constants will be specified later (see section 2.2). We can then rewrite the basic equations with dimensionless variables as follows:

\begin{equation}
\frac{\partial \sigma}{\partial \tau} = \frac{1}{4\pi x^3} \frac{\partial \dot{m}}{\partial x},
\label{basic1'}
\end{equation}

\begin{equation}
\dot{m} = 3\pi \frac{(x^{2}-x_{\rm H}^{2})^{\frac{\beta+2}{2}}}{\left[x^{2}-(1+\beta)x_{\rm H}^{2}\right]x^{\beta}} \frac{\partial}{\partial x} \left[\frac{x^{\beta+1}(x^{2}-\frac{3-\beta}{3}x_{\rm H}^{2})}{(x^{2}-x_{\rm H}^{2})^{\frac{\beta+2}{2}}} \mu \right].
\label{basic2'}
\end{equation}

\subsection{The N-shaped Equilibrium Curve}

To solve the basic equations (\ref{basic1'}) and (\ref{basic2'}), we need to prescribe the functional form of $\mu = \mu(\sigma)$.
This can be done by using the thermal equilibrium curve, on which heating and cooling are balanced.
The thermal equilibrium curve of the hyperaccretion flow has a characteristic N-shape, as is shown in the left panel of Figure 1 taken from \citet{kawanaka+13b}. They adopted $\alpha$ viscosity; 

\begin{equation}
\nu = \frac{2}{3} \alpha c_{\rm s} H
\label{alpha-viscosity}
\end{equation}
with $\alpha=0.1$, where $H$ is the half thickness of the disk and $c_{\rm s}$ is the speed of sound.
The vertical axis is $\dot{M}$ (mass accretion rate) and the horizontal axis is $\Sigma$ (surface density) in the left panel of Figure 1.
We model this equilibrium curve in a simplified form as is shown in the right panel of Figure 1:

For $\sigma < \sigma_{\rm A}$
\begin{equation}
\mu = x\sigma
\label{N-curve-A}
\end{equation}

For $\sigma_{\rm A} \leq \sigma < \sigma_{\rm B}$
\begin{eqnarray}
\mu =
\left\{
\begin{array}{ll}
\mu_{\rm A}\left(\frac{\displaystyle\sigma}{\displaystyle\sigma_{\rm A}}\right)^{(4x-2.0)} & (x \leq 0.5)\\
\mu_{\rm A}\left(\frac{\displaystyle\sigma}{\displaystyle\sigma_{\rm A}}\right)^{(2x-1.0)} & (x > 0.5)
\end{array}
\right.
\label{N-curve-B}
\end{eqnarray}

For $\sigma \geq \sigma_{\rm B}$
\begin{equation}
\mu = \mu_{\rm B}\frac{\sigma}{\sigma_{\rm B}}
\label{N-curve-C}
\end{equation}
where $\sigma_{\rm A}$ and $\sigma_{\rm B}$ are the values of $\sigma$ at the two critical points, `A' and `B', respectively, where the slope of the thermal equilibrium curve changes;
\begin{eqnarray}
\sigma_{\rm A} = 5.0 + \frac{10.0}{0.75}(x - 0.25), \quad \mbox{\rm and} \quad \sigma_{\rm B} = 15.0
\label{jyouken}
\end{eqnarray}
Likewise, $\mu_{\rm A}$ and $\mu_{\rm B}$ are the values of $\mu$ at the critical points, `A' and `B', respectively, and their explicit forms are calculated by using equations (\ref{N-curve-A}) and (\ref{N-curve-B}) as functions of $x$.

\begin{figure*}
\begin{minipage}{0.5\hsize}
\label{model-kawanaka}
%\vspace{-2.5cm}
\begin{center}
\FigureFile(80mm, 50mm){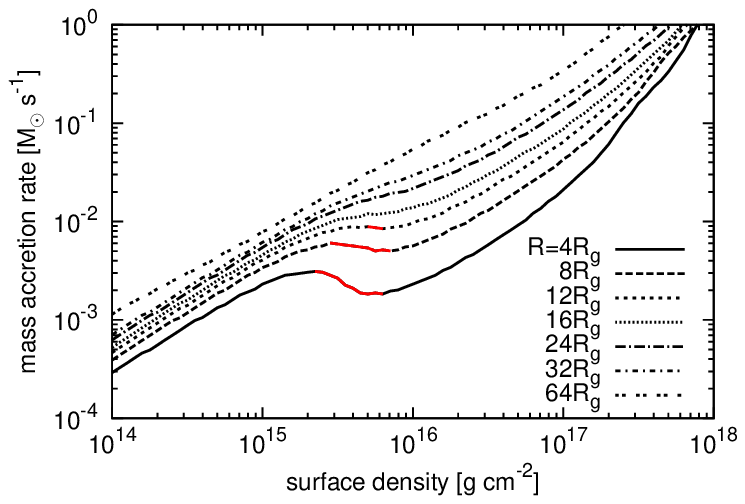}
\end{center}
\end{minipage}
\begin{minipage}{0.5\hsize}
\label{N-model}
\begin{center}
\FigureFile(80mm, 50mm){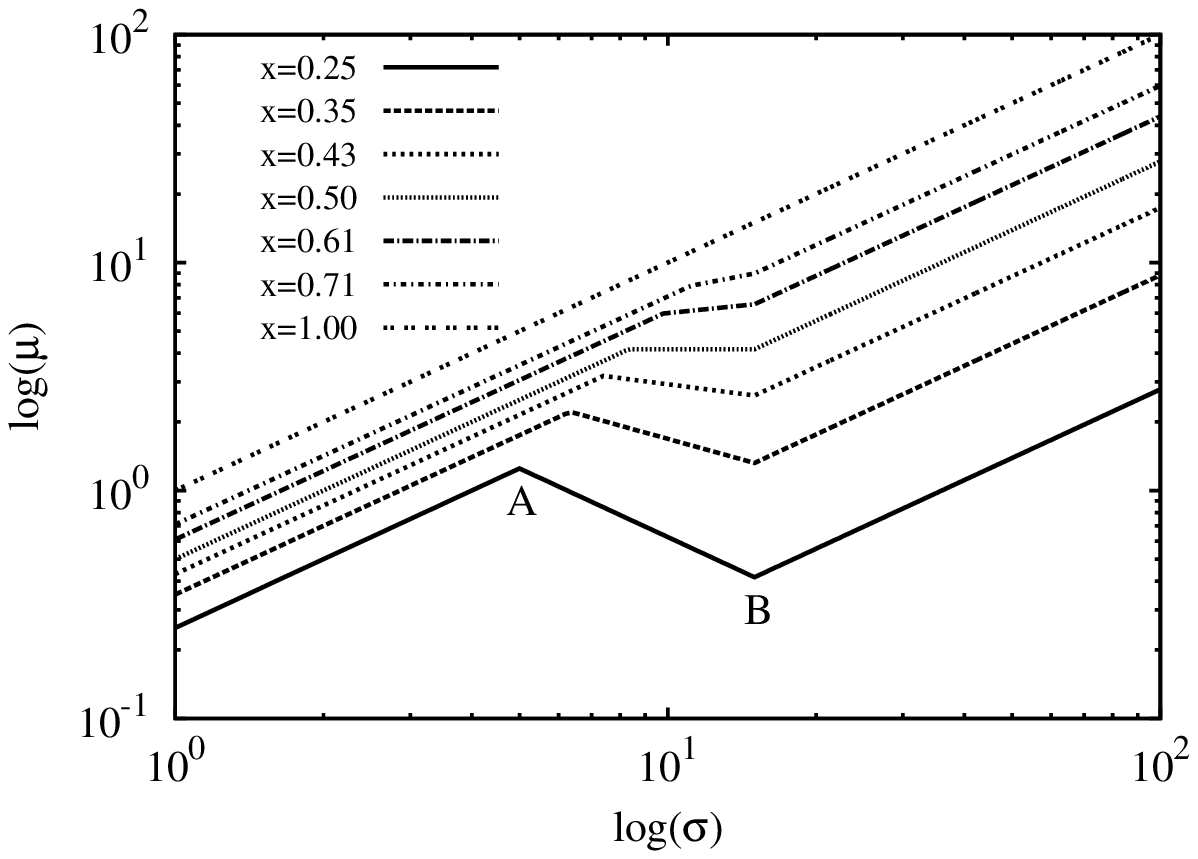}
\end{center}
%\vspace{1cm}
\end{minipage}
\caption{Two thermal equilibrium curves: The left panel (a) the curve calculated by \citet{kawanaka+13b}. The right panel (b) the simplified curve used in the present study. The symbols `A' and `B' denote the two critical points.}
\end{figure*}

Equations (\ref{N-curve-A})--(\ref{N-curve-C}) represent the lower, middle, and upper branches of the thermal equilibrium curve, respectively.
The principle for constructing this formula is as follows:
\begin{enumerate}
\item The value of the kinematic viscosity is constant (i.e., $\mu \propto \sigma$) on the lower and upper branches, while $\mu \propto 1/\sigma$ on the middle branch in the innermost region. The bigger the radius is, the smaller is the slope of the middle branch.
\item The entire equilibrium curve up-shifts as $x$ increases. In other words, the bigger $x$ is, the bigger is the value of $\mu$ for a fixed $\sigma$.
\item The N-shape is more enhanced at smaller radii, whereas the unstable branch (with a negative slope) disappears at $x \geq 0.5$.
\end{enumerate}

In the present study, we set $r_{\rm out} = 64R_{\rm g}$.
Here, $R_{\rm g} ( = GM_{\rm BH}/c^{2})$ is the gravitational radius.
By comparing Figure 1(a) with Figure 1(b), we assign the following values to $\dot{M}_{0}$ and $\Sigma_{0}$: 
\begin{equation}
\dot{M}_{0} = 3.5 \times 10^{-3}[M_{\solar}~{\rm s}^{-1}],
\label{Mdot0}
\end{equation}
\begin{equation}
\Sigma_{0} = 5.5 \times 10^{14}[{\rm g}~{\rm cm}^{-2}],
\label{Sigma0}
\end{equation}
for the black hole mass of $M_{\rm BH} = 3M_{\solar}$.
Inserting these normalization constants into equation (9) and simply assuming that $t_{0} \propto M_{\rm BH}$, we find
\begin{equation}
t_{0}(M_{\rm BH}) = 6.4 \times 10^{-2}\left(\frac{M_{\rm BH}}{3M_{\solar}}\right)[\rm s].
\label{t0}
\end{equation}

\subsection{Simplified Energy Equation}

As long as the system is in the thermal equilibrium, it evolves along the equilibrium curve.
It starts to deviate from the curve, however, once the instability takes place.
In order to describe the disk evolution in the state out of equilibrium, we introduce the following simple form of energy equation:

\begin{equation}
\frac{\partial \mu}{\partial \tau} = -\frac{\mu - \mu_{\rm eq}}{\tau_{\rm th}}
\label{thermal.eq}
\end{equation}
where $\mu_{\rm eq}$ is the value of $\mu$ on the thermal equilibrium curve for a given $\sigma$, and $\tau_{\rm th}$ represents the thermal timescale;
\begin{eqnarray}
\tau_{\rm th} = f \cdot \tau_{\rm vis} \quad \mbox{with} \quad \tau_{\rm vis} \equiv \frac{r^2}{\nu} = t_{0} \frac{\sigma}{\mu} x^{4}
\label{tau-th}
\end{eqnarray}
Here, $\tau_{\rm vis}$ represents the diffusion timescale at radius $r$ and is shorter at smaller $r$, and $f$ is a constant set to be $f = 0.1$ in the present study.

\subsection{Boundary Conditions and Initial Conditions}

There are two boundary conditions.
One is the choice of $\dot{m}_{\rm inj}$, mass injection rate at the outer edge of the disk (the outer boundary condition) and the other is the zero torque condition at the inner edge of the disk (the inner boundary condition);
\begin{eqnarray}
\dot{m} = \dot{m}_{\rm inj}(\tau) \quad \mbox{at} \quad r=r_{\rm out}\\
\mu = 0 \quad \mbox{at} \quad r=r_{\rm in}
\label{torque0}
\end{eqnarray}

In this study, we examine the following three cases for the way of mass injection from the outer boundary:
\begin{equation}
\mbox{Model 1 (slow rise):}~\dot{m}_{\rm inj}=\dot{m}_{0}\exp (\tau/10)~\mbox{with}~\dot{m}_{0} = 3\pi \times 8.60
\label{mdot-inj-1}
\end{equation}
\begin{equation}
\mbox{Model 2 (rise):} \quad \dot{m}_{\rm inj}=\dot{m}_{0}\exp(\tau) \quad \mbox{with} \quad \dot{m}_{0} = 3\pi \times 8.60
\label{mdot-inj-2}
\end{equation}
\begin{equation}
\mbox{Model 3 (decay):} \quad \dot{m}_{\rm inj}=\dot{m}_{0}\exp(-\tau) \quad \mbox{with} \quad \dot{m}_{0} = 3\pi \times 33.00
\label{mdot-inj-3}
\end{equation}
Note that the value of $\dot{m}_{0}$ of Models 1 and 2 (or Model 3) has been chosen in order that the whole disk can be on the lower (upper) branch; i.e., $\dot{m}_{\rm inj} < 3\pi \times 8.69$ ($\dot{m}_{\rm inj} > 3\pi \times 32.65$).
Also note that the rise timescale of Model 1 is $10~t_{0} = 0.64 (M_{\rm BH} / 3M_{\solar})$ [{\rm s}] [see equation (\ref{t0})].

The initial condition is a steady disk corresponding to $\dot{m}_{0}$.
By integrating equation (\ref{basic2'}) with respect to $x$, we obtain
\begin{equation}
\mu = \frac{\dot{m}}{3\pi}\frac{(x^{2}-x_{\rm H}^{2})^{\frac{\beta+2}{2}}}{\left[x^{2}-(\frac{3-\beta}{3})x_{\rm H}^{2}\right] x^{\beta+1}} \left[\frac{x^{\beta+1}}{(x^{2}-x_{\rm H}^{2})^{\frac{\beta}{2}}} - \frac{x_{\rm in}^{\beta+1}}{(x_{\rm in}^{2}-x_{\rm H}^{2})^{\frac{\beta}{2}}} \right],
\label{steady-mu}
\end{equation}
and the initial value of $\mu$ is calculated by inserting $\dot{m}_{0}$ into the equation above.
We can then solve time-dependent equations (\ref{N-curve-A})--(\ref{N-curve-C}) to obtain surface density evolution at each $x$.
Actual numerical procedure is explained in Appendix 1.

We set the inner boundary at $r_{\rm in} = 3.84R_{\rm g}$ or $x_{\rm in} = 0.24$.
The inner radius corresponds to the innermost stable circular orbit of a black hole with the dimensionless spin parameter $a_{*}$ of 0.6, for which $r_{\rm H} = 1.8R_{\rm g}$.
We finally get $\beta = (r_{\rm ms}/r_{\rm H})-1 \approx 1.12$.

\section{Results}

\subsection{Basic Consideration Based on a Steady Model}

How does a transition from the lower branch to the upper branch proceed in the disk when the mass injection rate gradually increases?
To answer to this question, we first show in Figures 2(a) and 3(a) a series of the surface-density distribution expected to be the steady disk model for several mass injection rates; $\dot{m}_{\rm inj} = \dot{m}_{0} \exp(\tau/10)$ with $\tau = 0.2,~0.8,~1.4$, and $2.0$ in Figure 2(a) while $\tau = 0.0,~0.3,~\cdots,~3.0$ in Figure 3(a), respectively [recall equation (\ref{mdot-inj-1})].
We, here, assume that the disk is on the lower ($\sigma$) branch of the N-shaped curve as long as $\sigma < \sigma_{\rm A}$ but that it is on the upper branch otherwise.
Note that each curve in Figure 3(a) except for the lowest one is vertically offset by the value depending on the elapsed time.

In these plots the mass accretion rates are taken to be the same everywhere by the assumption that every portion of the disk can immediately adjust to any changes in $\dot{m}_{\rm inj}$ given at the outer boundary.
As $\dot{m}_{\rm inj}$ increases, $\dot{m}$ at each radius gradually increases, so does the value of $\sigma$.
Eventually a transition from the lower ADAF branch to the upper NDAF branch takes place when $\sigma > \sigma_{\rm A}$.
We can see in Figures 2(a) and 3(a) that the surface density rapidly increases at the radius where an upward transition occurs.
After the transition, neutrino cooling is efficient.
Hence, neutrino is emitted from the high-$\sigma$ regions.

The steady model is simple and easy to calculate, since it requires no numerical technique, however, there exist certainly limitations. For example, it should take time for the influence of varying mass injection rates to prevail over the disk plane, but such a time delay is totally neglected in the steady model.
Also an upward transition cannot occur in an infinitesimal time but a transition is assumed to be instantaneous in the steady model.
Further, mass inflow into and/or outflow from the transition region, which should be necessary to make a local change in $\sigma$, are not properly taken into account.
We thus need to perform numerical simulations.

\subsection{Model 1 (slow rise)}

\subsubsection{Surface-Density Evolution}

The first case is Model 1, in which the mass injection rate slowly increases.
Figures 2(b) and 3(b) display the global evolution of the surface-density distribution in slightly different ways.
The expansion of the high-$\sigma$ region, where neutrino emission is significant, is clear in these figures.
The upward transition to the upper (NDAF) branch is indicated by a rapid increase of $\sigma$.

\begin{figure*}
\begin{minipage}{0.5\hsize}
\label{steady-sigma}
%\vspace{-1cm}
\begin{center}
\FigureFile(80mm, 50mm){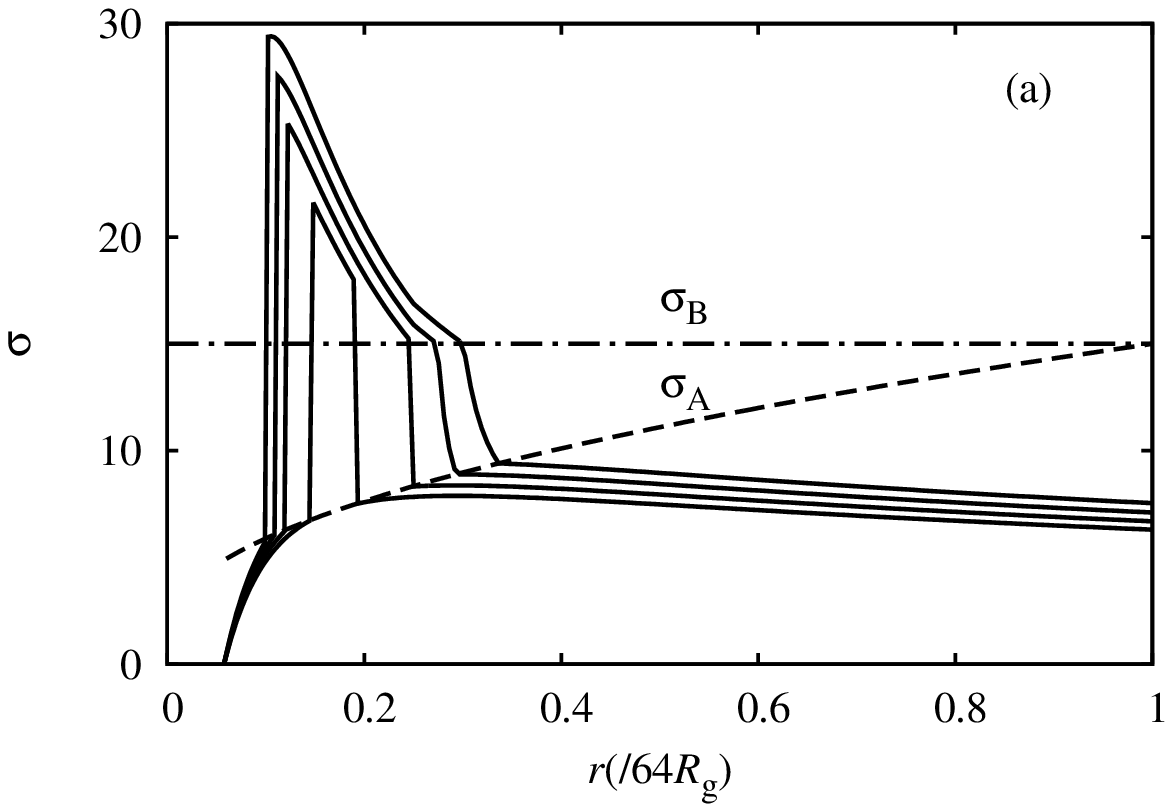}
\end{center}
\end{minipage}
\begin{minipage}{0.5\hsize}
\label{sigma-hikaku}
\begin{center}
\FigureFile(80mm, 50mm){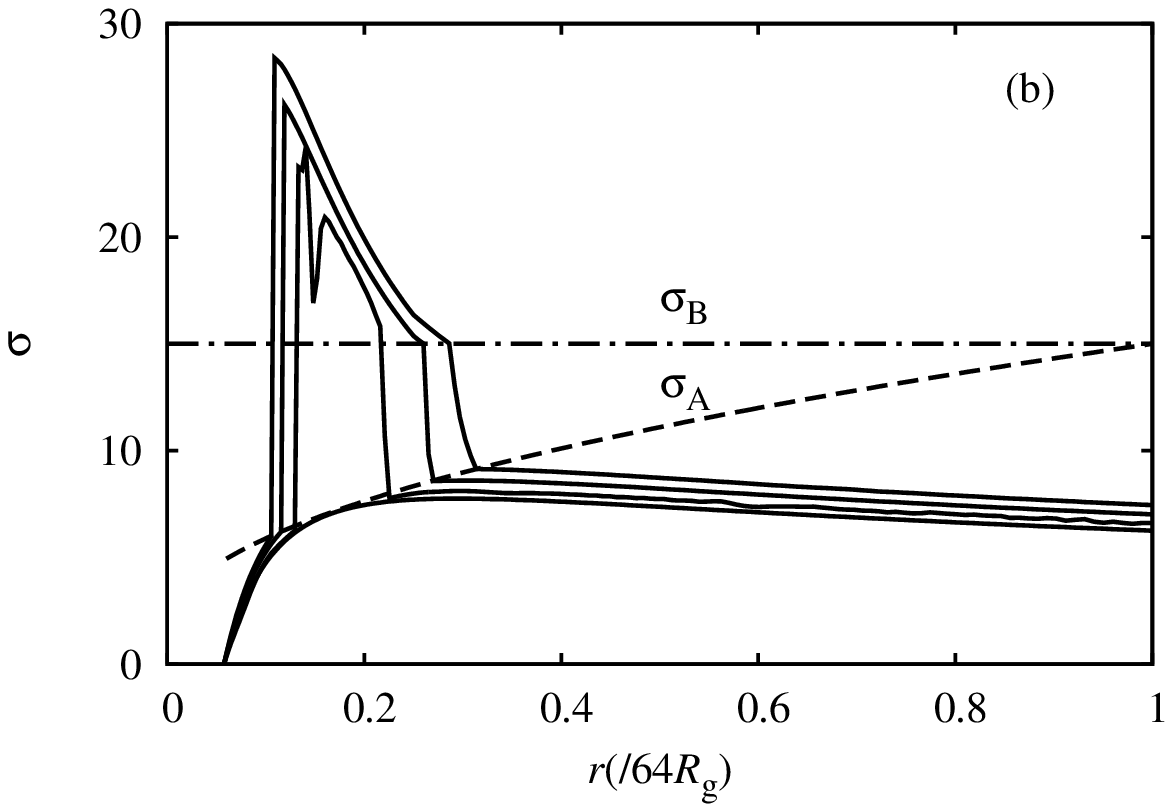}
\end{center}
%\vspace{1cm}
\end{minipage}
\caption{Variety of surface-density ($\sigma$) distributions based on the steady model (a) and on the time-dependent model (b) for various mass injection rates of $\dot{m}_{\rm inj} = \dot{m}_{0} \exp(\tau/10)$ with $\tau = 0.2,~0.8,~1.4$, and $2.0$. The critical lines ($\sigma_{\rm A}$ and $\sigma_{\rm B}$) are also plotted by the dashed line and the dash-dot line, respectively.}
\end{figure*}

\begin{figure*}
\begin{minipage}{0.5\hsize}
\label{steady1}
%\vspace{-1cm}
\begin{center}
\FigureFile(80mm, 50mm){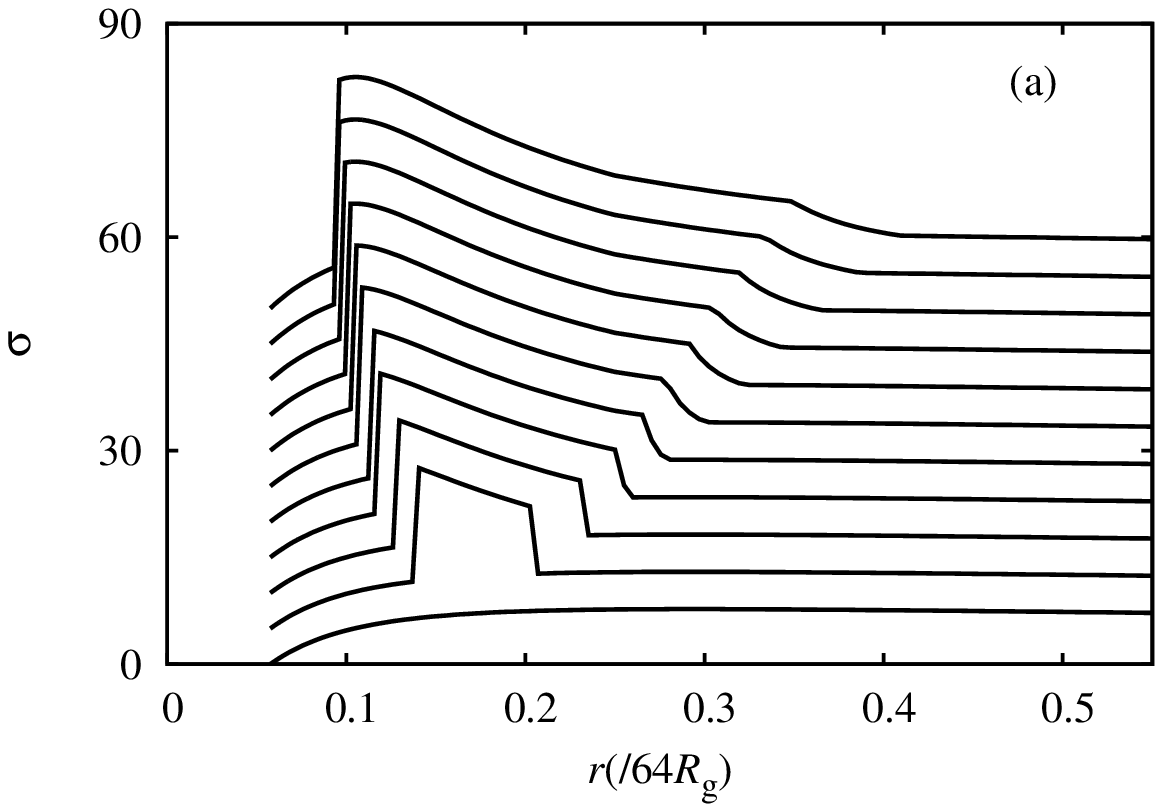}
\end{center}
\end{minipage}
\begin{minipage}{0.5\hsize}
\label{sigma1}
\begin{center}
\FigureFile(80mm, 50mm){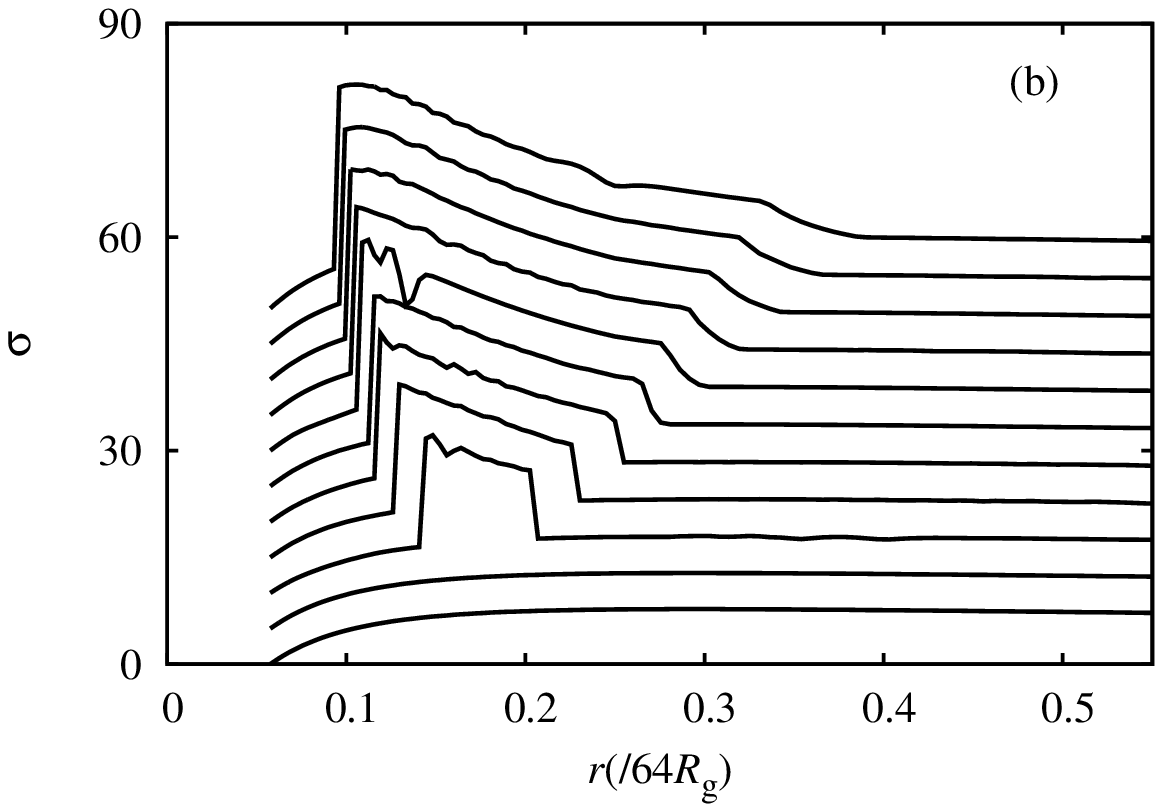}
\end{center}
%\vspace{1cm}
\end{minipage}
\caption{Series of $\sigma$ distributions in the steady model (panel a) and in Model 1 calculations (panel b). The corresponding mass injection rates are $\dot{m}_{\rm inj} = \dot{m}_{0} \exp(\tau/10)$ with $\tau$ = 0.0, 0.3, $\cdots$, and 3.0. Each line is offset vertically by the value of $50/3 \times \tau$ for visible clarify.}
\end{figure*}

Let us carefully examine to what extent the steady model depicted in Figure 2(a) reproduces the time-dependent calculations shown in Figure 2(b).
In the steady model, the mass accretion rate ($\dot{m}$) is everywhere equal to the mass injection rate ($\dot{m}_{\rm inj}$); in other words, the whole disk is assumed to immediately respond to a change of $\dot{m}_{\rm inj}$ at the outer boundary.
Another big assumption in the steady model is that a transition from one branch to another at a certain radius does not affect its environment.
From these two panels, we notice that the time of an upward transition at a certain radius in the time-dependent model is delayed, compared with that at the same radius in the steady model.
This feature can also be confirmed by comparing Figures 3(a) and (b) and is obviously the effect of the finite accretion timescale.
We also find that the $\sigma$ profile shows fluctuating patterns.
These fluctuating patterns are more clearly displayed in Figure 3(b).
This is caused by rapid mass exchanges between the neighboring regions associated with branch transitions.
Hereafter we call such fluctuations as `non-steady' effects. Obviously they are induced by `non-steady' mass accretion flow (i.e., locally enhanced mass flow). We will discuss their physical meaning in details in the next subsection.\\

\begin{figure}
%\vspace{-1cm}
\begin{center}
\FigureFile(80mm, 50mm){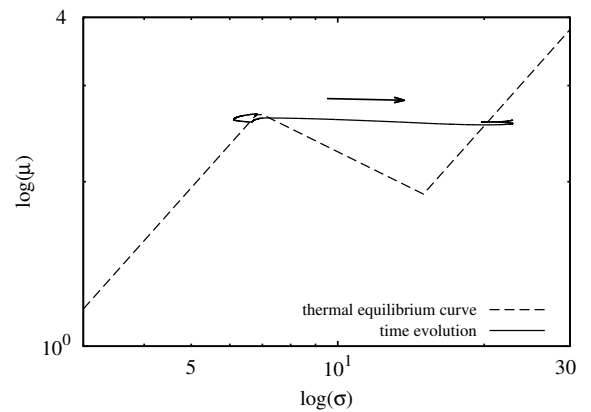}
\end{center}
%\vspace{1cm}
\caption{The actual behavior on the $\sigma$-$\mu$ plane at the radius $x=0.39$, where an upward transition is initiated at $\tau \leq 0.357$: The solid line represents the time evolutionary path, while the dashed line represents the thermal equilibrium curve. The arrow represents the direction of the time evolution.}
\label{nusigma78}
\end{figure}

\subsubsection{Mechanism of Transitions Induced by Non-steady Effects}

In order to understand the non-steady effects, more careful analysis is necessary regarding the behavior of each part of the disk.
This can be done by using the $\sigma$-$\mu$ plane at fixed radii. 
A typical example is displayed in Figure \ref{nusigma78}.
We see that the evolutionary path shown by the solid line grossly deviates from the thermal equilibrium curve indicated by the dashed line.
We can understand this fact in the following way.

Suppose that the critical point A is reached at some radius.
Although there are no longer low branch solutions above that point, $\sigma$ should steadily increase, since mass is continuously supplied from the neighboring outer zone.
Thus, the disk should evolve in the right direction on the $\sigma$-$\mu$ plane.
This is exactly what we see in Figure \ref{nusigma78}.

After reaching the upper branch, however, unexpected behavior appears. That is, the evolutionary path fluctuates around the equilibrium point, on the upper branch.
This is a result of complex mass exchanges between the neighboring sites.
To understand how this occurs, we make the following thought experiments.

At first we wish to note that the exact behavior of the disk on the $\sigma$-$\mu$ plane depends critically on the ratio of the thermal timescale ($\tau_{\rm th}$, on which $\mu$ changes) to the viscous timescale ($\tau_{\rm vis}$, on which $\sigma$ changes).
Let us, hence, consider two hypothetical cases: Case A transition in which $\tau_{\rm th}/\tau_{\rm vis} \ll 1$ as is expected in a standard-type disk (see, e.g., Kato et al. 2008, chapter 3) and Case B transition in which $\tau_{\rm th}/\tau_{\rm vis} \sim 1$. (The actual simulation will be shown later.)

\begin{figure*}
\begin{minipage}{0.99\hsize}
\label{mangaA}
\begin{center}
\FigureFile(150mm, 50mm){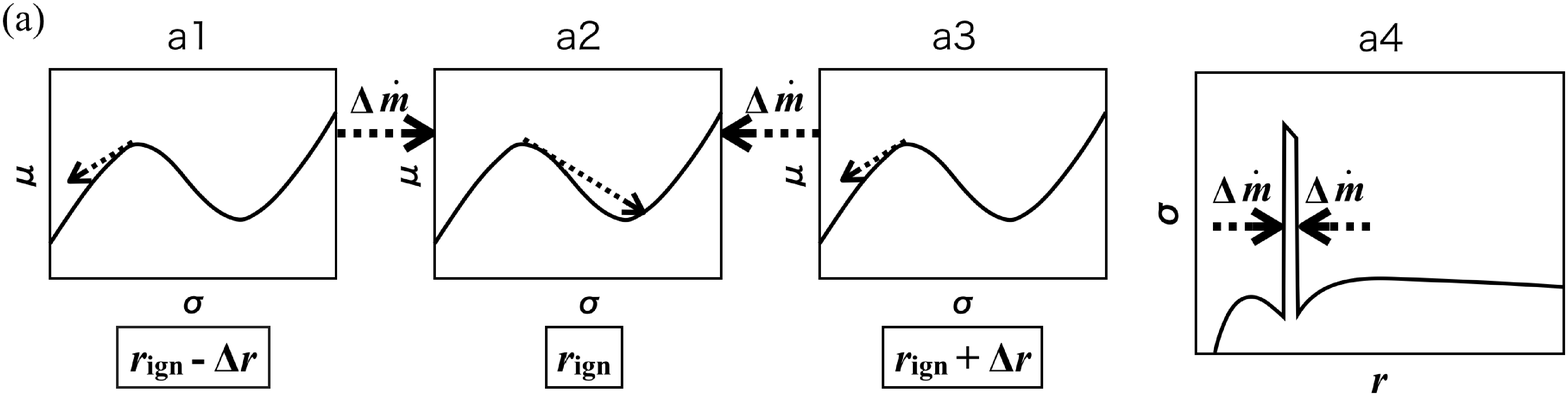}
\end{center}
\end{minipage}
\\
\begin{minipage}{0.99\hsize}
\label{mangaB}
\begin{center}
\FigureFile(150mm, 50mm){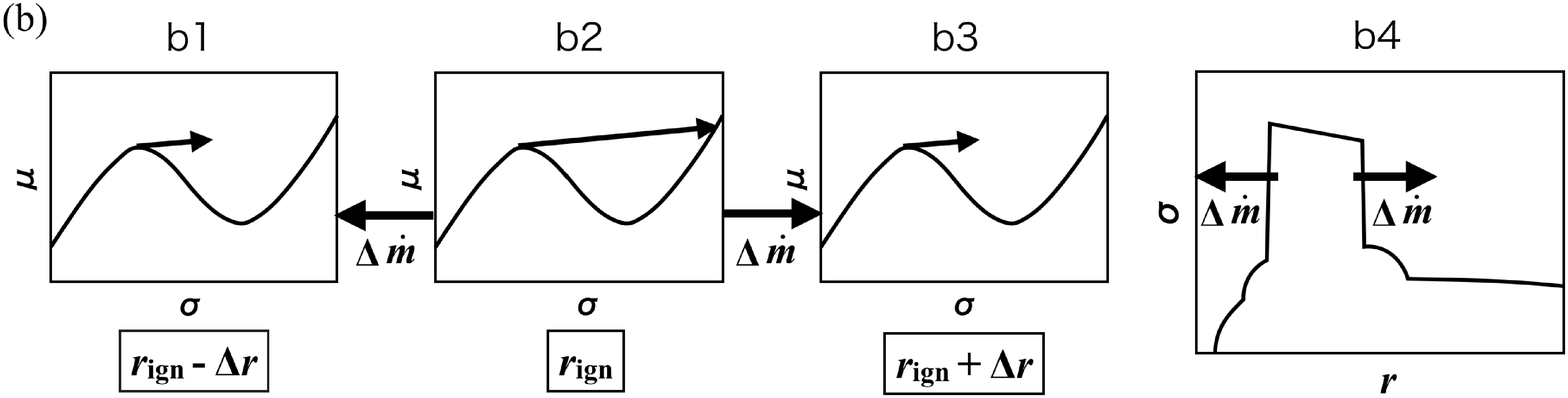}
\end{center}
\end{minipage}
\\
\begin{minipage}{0.99\hsize}
\label{mangaC}
\begin{center}
\FigureFile(150mm, 50mm){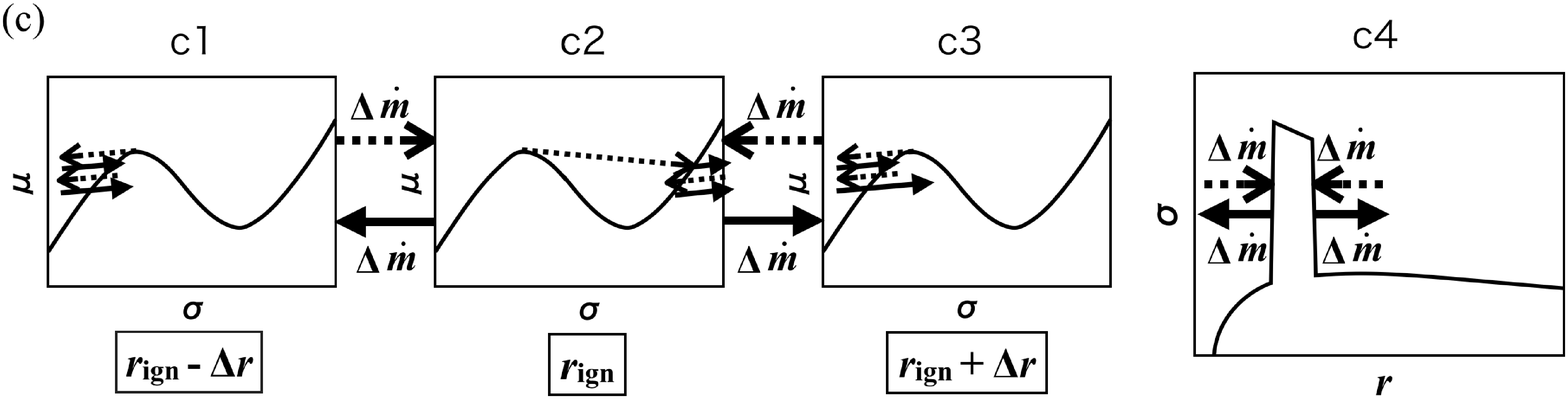}
\end{center}
\end{minipage}
\caption{Schematic pictures explaining how a transition occurs by non-steady effects. Here, we consider two hypothetical cases in the upper and middle panels and one actual calculation results in the bottom panels. Left nine panels illustrate time-dependent evolution of the disk on the $\sigma$-$\mu$ plane and associated mass flow at the ignition point $r_{\rm ign}$ (panels a2, b2, and c2) and at its neighboring sites (other panels). Note that $\Delta \dot{m}$ (non-steady mass flow) represents mass flow other than the global mass flow (see text for the definition). Right three panels a4, b4, and c4 illustrate the time-dependent evolution of the disk on the $r$-$\sigma$ plane.}
\label{manga-1}
\end{figure*}

In Case A transition (see the upper four panels of Figure \ref{manga-1}), we assume $\tau_{\rm th} / \tau_{\rm vis} \ll 1$. A short thermal timescale means that a system can quickly adjust to any changes added externally so that it should always stay around the thermal equilibrium state. [We assume that the disk is thermally stable (see, e.g., \cite{kawanakamineshige07}), as in the case that we are discussing.]
Let us denote by $r_{\rm ign}$ the ignition radius, where the instability sets out.
Since the disk should evolve along the thermal equilibrium curve, $\sigma$ should increase accordingly, while $\mu$ should decrease during an upward transition at $r_{\rm ign}$ (see panel a2 of Figure \ref{manga-1}).
A non-steady mass flow ($\Delta \dot{m}$) is then induced from its neighboring sites (with high $\mu$ values) to the ignition region (with low $\mu$ values) as is illustrated in panel a4, since crudely we have $\dot{m} \propto \partial (x\mu) / \partial x$ from equation (\ref{basic2'}).
Here, the non-steady mass flow is defined by

\begin{equation}
\Delta \dot{m}(r) \equiv \dot{m} - \langle \dot{m} \rangle,
\label{non-local-flow}
\end{equation}
where the averaged value, $\langle \dot{m} \rangle$, is the mass flow rate averaged over a region around the radius, $r$.
This non-steady mass flow assures an increase of $\sigma$ at the ignition point, whereas $\sigma$ in the neighboring sites should decrease (see panels a1 and a3 of Figure \ref{manga-1}).
To conclude, a transition at a certain point does not trigger further transitions in the surrounding region.

In order for a transition to propagate into the neighboring regions, these regions should acquire mass from the ignition site (at $r_{\rm ign}$).
For this to occur, the value of $\mu$ has to increase at $r_{\rm ign}$, as is illustrated in panel b2 of Figure \ref{manga-1} (referred to as Case B transition); i.e., the disk evolutionary path on the $\sigma$-$\mu$ plane should largely deviate from the equilibrium curve.
For this sort of behavior to take place we should require $\tau_{\rm th}/\tau_{\rm vis} \sim 1$, but this never happens in a standard type disk with no instability, since roughly we estimate

\begin{equation}
\tau_{\rm th} / \tau_{\rm vis} \sim (H/r)^{2} \ll 1,
\label{standard-relation}
\end{equation}
(with $H$ being the scale-height of the disk), and since the aspect ratio is expected to be small, $(H/r) \ll 1$.
We wish to emphasize, however, that this inequality no longer holds, once an instability grows and produces abrupt, spatial variations of physical quantities.
Under such circumstances, equation (\ref{standard-relation}) should be written as
\begin{equation}
\tau_{\rm th} / \tau_{\rm vis} \sim (H/\Delta r)^{2} \sim 1,
\label{not-standard-relation}
\end{equation}
with $\Delta r$ being the spatial scale, on which the value of $\mu$ varies.
If this were the case, the evolution of $\sigma$ profile would be like that illustrated in panel b4 of Figure \ref{manga-1}.

Which is actually the case in the time-dependent calculation?
The resultant time evolution of an unstable disk is illustrated as Case C in the lower four panels of Figure \ref{manga-1}.
When a transition occurs, $\mu$ only slightly decreases at $r_{\rm ign}$ [see Figure \ref{nusigma78}].
This apparently looks like Case A except for one important difference; that is, the disk trajectory starts to oscillate around its equilibrium value through interactions with neighboring regions after the transition to the upper branch.
(The meaning of `interactions' will be discussed in more details in Sec. 3.2.3.)
Owing to these fluctuations, the values of $\mu$ in the surrounding regions also begin to oscillate around their equilibrium values, and such fluctuations in $\mu$ eventually induce mass accretion flow from or into the neighboring region, thereby occasionally triggering transitions [see Figure 5(c)].
These are purely non-steady effects.

Importantly, the onset of the instability at some radius does not always trigger the instability in adjacent regions. The instability zone is confined in a relatively narrow region. In other words, the instability does not promote coherent behavior of the entire disk. This point will be discussed in section 4.4 in comparison with the case of disks with the `S'-shaped equilibrium curve.\\

\subsubsection{Wave Propagation}

There is another interesting feature unique to unstable accretion disks; the influence of rapid variations in $\dot{m}$ propagates from $r_{\rm ign}$ in the radial direction, as is depicted in Figure 6(a).
We can see a wavy pattern in the $\dot{m}$ distribution propagating over the disk plane from $r_{\rm ign}$ both in the inner and outer directions.
Fluctuations in $\dot{m}$ drive small amplitude fluctuations in $\sigma$ [see equation (\ref{basic1'})].

\begin{figure*}
\begin{minipage}{0.5\hsize}
\label{mdot-wave-1}
%\vspace{-1cm}
\begin{center}
\FigureFile(80mm, 50mm){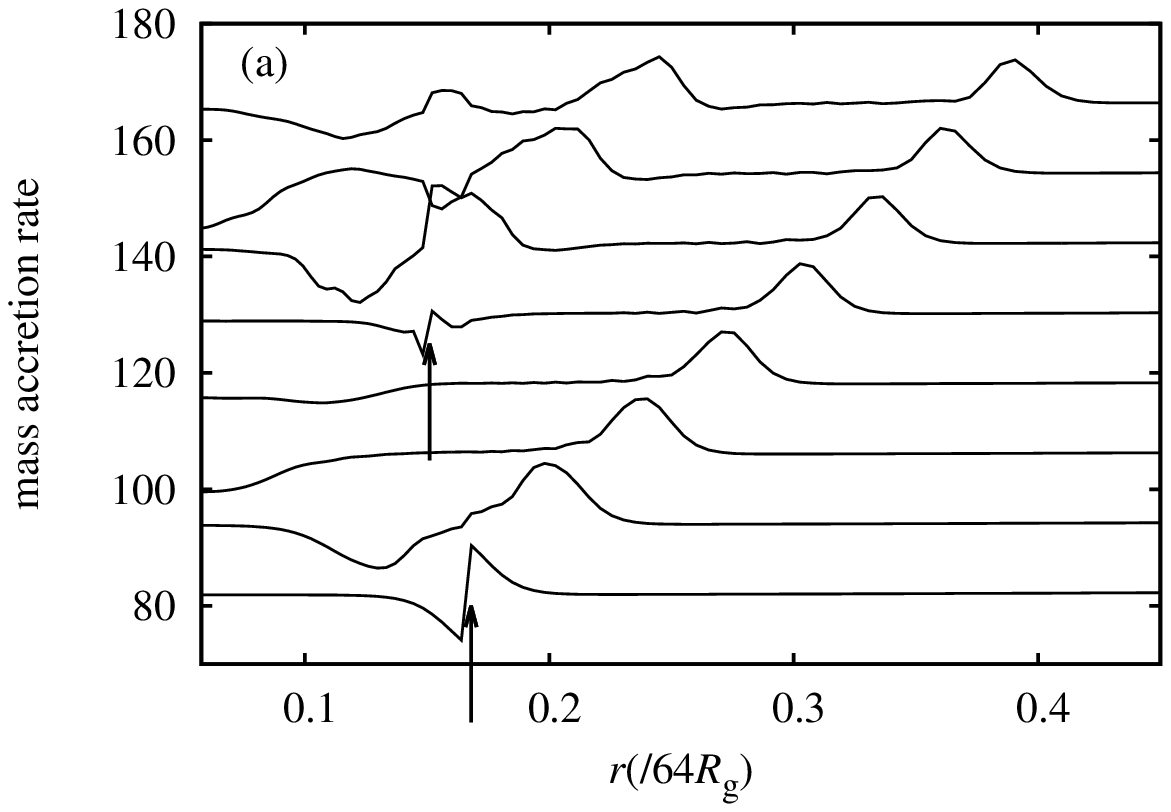}
\end{center}
\end{minipage}
\begin{minipage}{0.5\hsize}
\label{mdot-variation-1}
\begin{center}
\FigureFile(80mm, 50mm){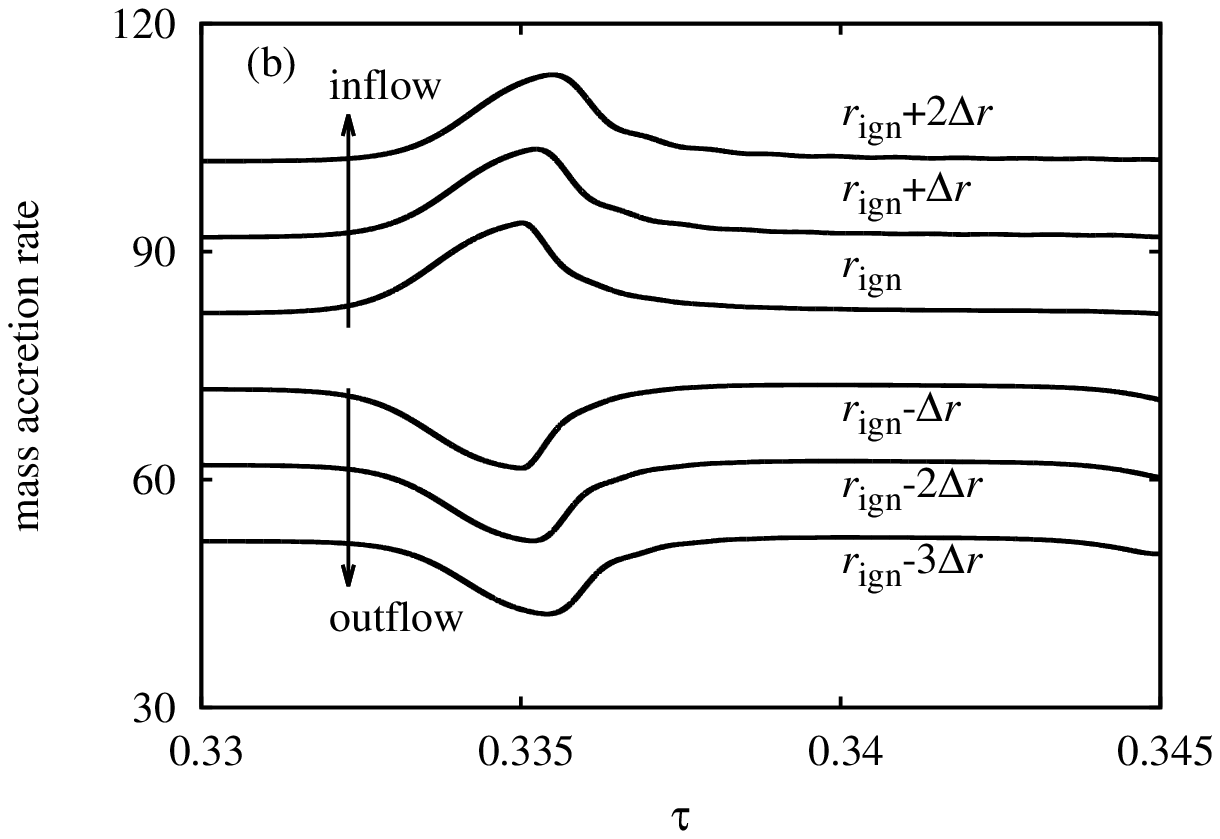}
\end{center}
%\vspace{1cm}
\end{minipage}
\caption{The vertical axis ``mass accretion rate'' represents $\dot{m}$. (a) Propagation of the $\dot{m}$ variation wave due to the instability over the disk plane. The elapsed times are 0.334, 0.337, 0.340, $\cdots$, and 0.355, and for visible clarity each curve is offset upward by 0, 12, 24, $\cdots$, and 84, respectively. The two arrows indicate the times and places of the onset of the instability. (b) Time variations of $\dot{m}$ at several radii illustrated on the $\tau$-$\dot{m}$ plane. Each curve is offset by $-30,~-20,~-10$, $\cdots$, and 20, respectively.}
\end{figure*}

Figure 6(b) illustrates the time developments of the accretion rate ($\dot{m}$) at several fixed radii. Here, by $\dot{m}$ at $r$ (= $r_{\rm ign}, r_{\rm ign} \pm \Delta r$, $\cdots$) we mean the mass flow from the radius $r + \Delta r$ to the radius $r$.
The upper three curves (or the lower three curves) represent non-steady mass inflow (outflow) towards the ignition point from the outer (inner) adjacent zones.
It is important to note time delays in the $\dot{m}$ response at the distant points (in comparison with that at the closer points) to the ignition radius, indicating that the fluctuations are not locally confined but spatially propagate and cause $\sigma$ variations in the neighboring sites. When the values of $\sigma$ exceed its critical value, $\sigma_{\rm A}$, an upward transition can be triggered.
The transition can occur, even when $\sigma$ is expected to be below $\sigma_{\rm A}$ in the steady model for a given $\dot{m}_{\rm inj}$.

Here, we discuss what is meant by ``interactions'' in Sec. 3.2.2. 
The unique features in Case C transitions reside in back-reactions of the non-steady mass flow ($\Delta {\dot m}$) [see the upper and lower panels of Figure 5].
According to our simulations, $\sigma$ and $\mu$ values change in the neighboring region after their receiving non-steady mass inflow. The changes in $\mu$ will then modify the mass flow pattern in a next moment, since ${\dot m} \propto \partial (x\mu) / \partial x$.
The modified mass flow pattern in the next moment produces further changes in $\sigma$ and $\mu$ values, which sometimes cause a negative feedback.
This is the back-reaction of the non-steady mass flow, which is not considered in Case A transition. The transition wave eventually damps but on longer timescale than the thermal time.
This explains why the solution oscillates around the equilibrium curve even after the thermal timescale.\\

\subsubsection{Neutrino Luminosity and Mass Accretion Rate}

The neutrino luminosity can, in principle, be calculated by

\begin{equation}
L_{\nu} = \int^{r_{\rm out}}_{r_{\rm in}} \frac{GM_{\rm BH}\dot{M}}{2r^{2-\beta}(r-r_{\rm H})^{\beta}} dr = L_{0} \int^{1}_{x_{\rm in}} \frac{\dot{m}}{64x^{3-2\beta}(x^{2}-x_{\rm H}^{2})^{\beta}} dx,
\label{lumi-neu}
\end{equation}
with
\begin{equation}
L_{0} = \frac{GM_{\rm BH}\dot{M}_{0}}{R_{\rm g}}.
\label{L0}
\end{equation}
Cautions should be taken here, since $\dot{m}$ in equation (\ref{lumi-neu}) should be mass flow rate on the upper (NDAF) branch.
We, hence, need special arrangements when calculating neutrino luminosity from the part on the lower branch, which has very little contribution to the neutrino luminosity, even if $\dot{m}$ is relatively large.
In actual calculations, therefore, we use equation (\ref{N-curve-C}) even when $\sigma < \sigma_{\rm B}$ to calculate $\mu$ from $\sigma$, and then calculate $\dot{m}$ from equation (\ref{steady-mu}) by assuming the steady-disk relation.

The resultant neutrino light curve is shown in Figure 7(a). In the same figure, we plot two more light curves representing the cases with no N-shaped equilibrium curves nor instability.
The upper (or lower) straight line labeled with `steady 1'(`steady 2') is the light curve that we would expect if the lower (upper) branch would extend up (down) to high (low) $\sigma$ regimes.

Comparing these light curves, we find two key features.
First, the overall shape of the light curve is steeper than the cases with no instability. The main reason for this resides in the `N'-shape; that is, the value of $\sigma$ and, hence, the $\mu$-value suddenly increases on a shorter timescale than the viscous timescale (on which $\dot{m}_{\rm inj}$ increases) after the onset of the instability.

The second key feature of the light curve is that the luminosity abruptly increases every time a transition takes place at a certain radius, resulting in step-function-like variations.
This is because that the instability is confined in a narrow region and does not produce coherent variations over a wide range of the disk plane, as mentioned above.

We compare with the mass accretion rate at radius $r_{\rm in}$ with the mass injection rate in Figure 7(b).
We can see significant variability in the mass accretion rate.
This may produce rapid variations in the electromagnetic radiation via the Blandford-Znajek process (to be discussed in section 4.3).

\begin{figure*}
\begin{minipage}{0.5\hsize}
\label{case1-lumi-neu}
%\vspace{-1cm}
\begin{center}
\FigureFile(80mm, 50mm){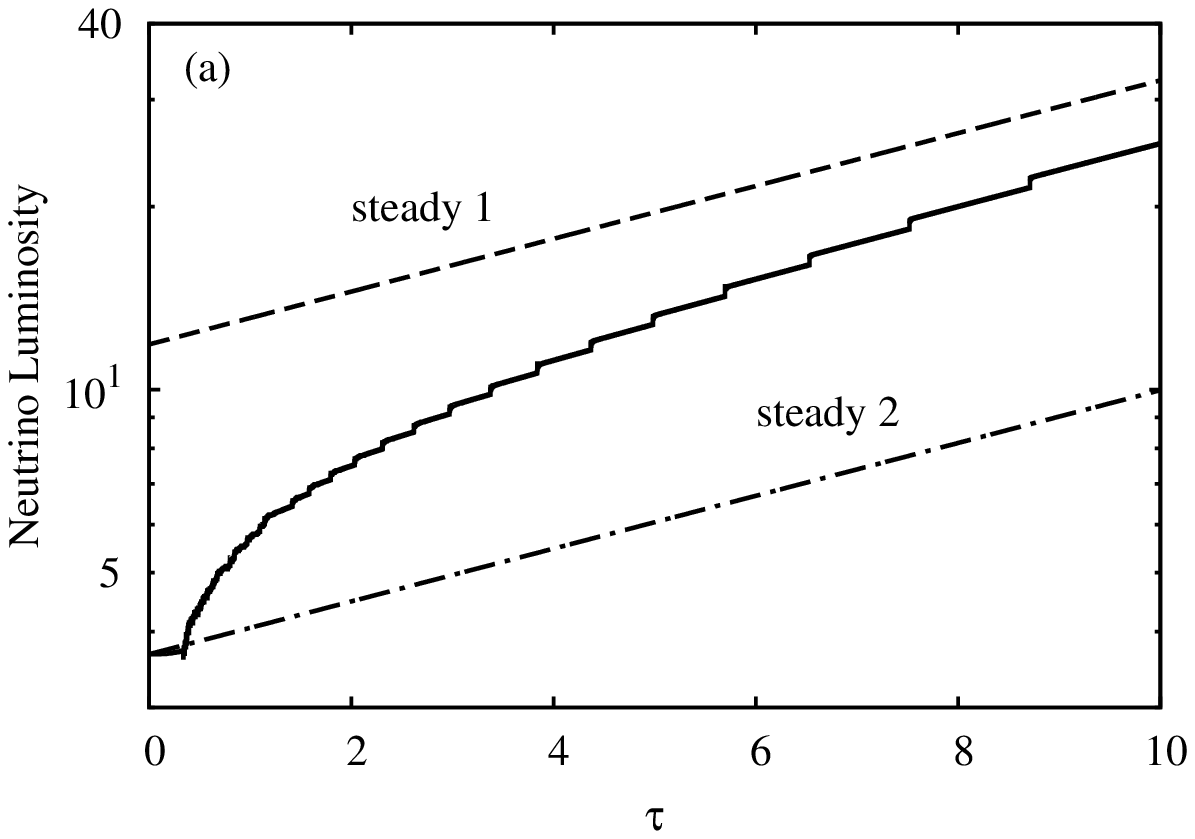}
\end{center}
\end{minipage}
\begin{minipage}{0.5\hsize}
\label{mdot-acc-1}
\begin{center}
\FigureFile(80mm, 50mm){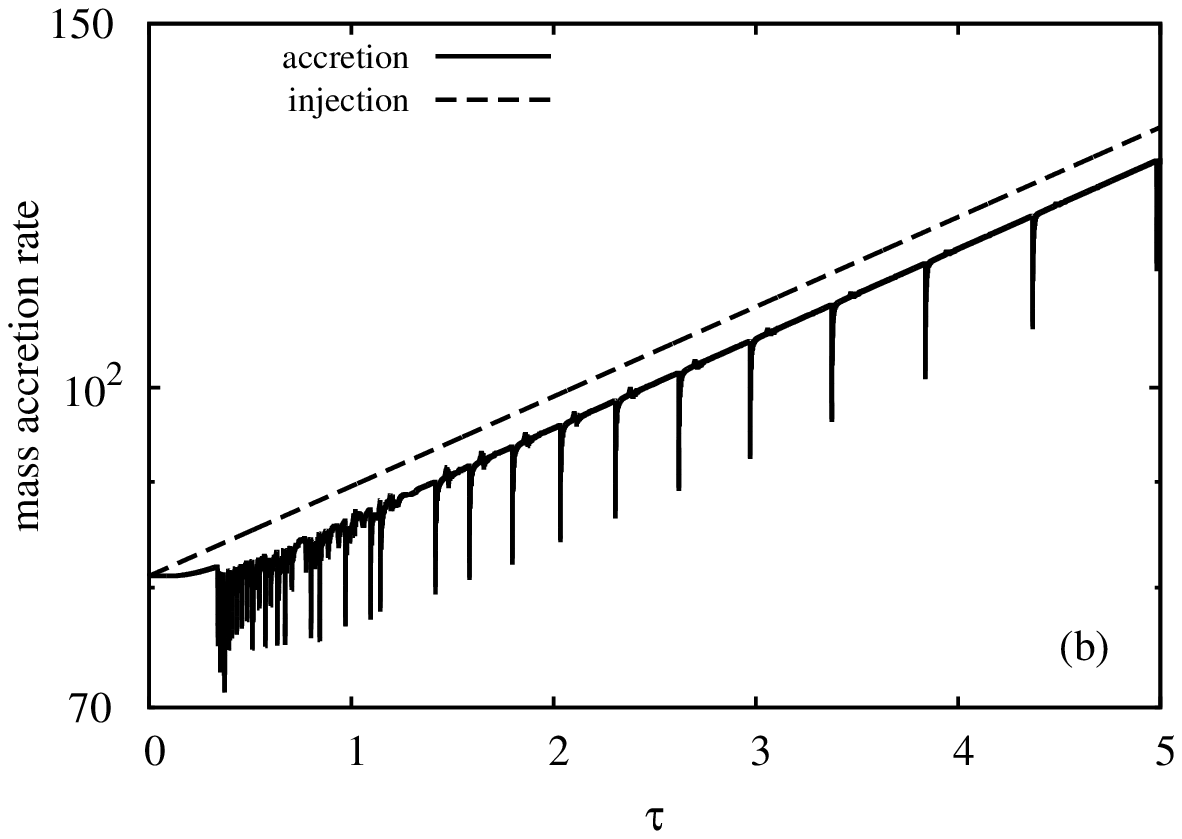}
\end{center}
%\vspace{1cm}
\end{minipage}
\caption{(a) The solid line represents the calculated neutrino luminosity in Model 1. The two dashed lines (`steady 1' and `steady 2') show what the light curves would be in case that the thermal equilibrium curve were not N-shaped; that is, we assume in `steady 1' that the lower branch [equation (15)] would extend up to the higher $\sigma$ regime while in `steady 2' the upper branch [equation (17)] would extend down to the lower $\sigma$ regime.
(b) Time variations of mass accretion rate at the inner edge of the disk (solid line) and of the mass injection rate (dashed line), respectively.}
\end{figure*}

\subsection{Model 2 (rise)}

The second case is Model 2, in which the mass injection rate increases more rapidly. We show the global evolution of the surface-density distribution in Figure 8(a). 
We can see that the region on the upper (NDAF) branch expands from $r_{\rm ign}$ both in the inner and the outer directions more rapidly than in Model 1.
This is simply because of more rapid increase in $\dot{m}_{\rm inj}$ in Model 2 than in Model 1.
If we have a closer look at this figure, moreover, we see that the out-going transition wave propagates faster than the in-going wave.

We show in Figure 8(b) the neutrino luminosity in Model 2. We can see that the luminosity rises more smoothly in Model 2 than in Model 1 because the speed of the instability propagation is faster than in Model 1.

\begin{figure*}
\begin{minipage}{0.5\hsize}
\label{sigma-rapid}
%\vspace{-1cm}
\begin{center}
\FigureFile(80mm, 50mm){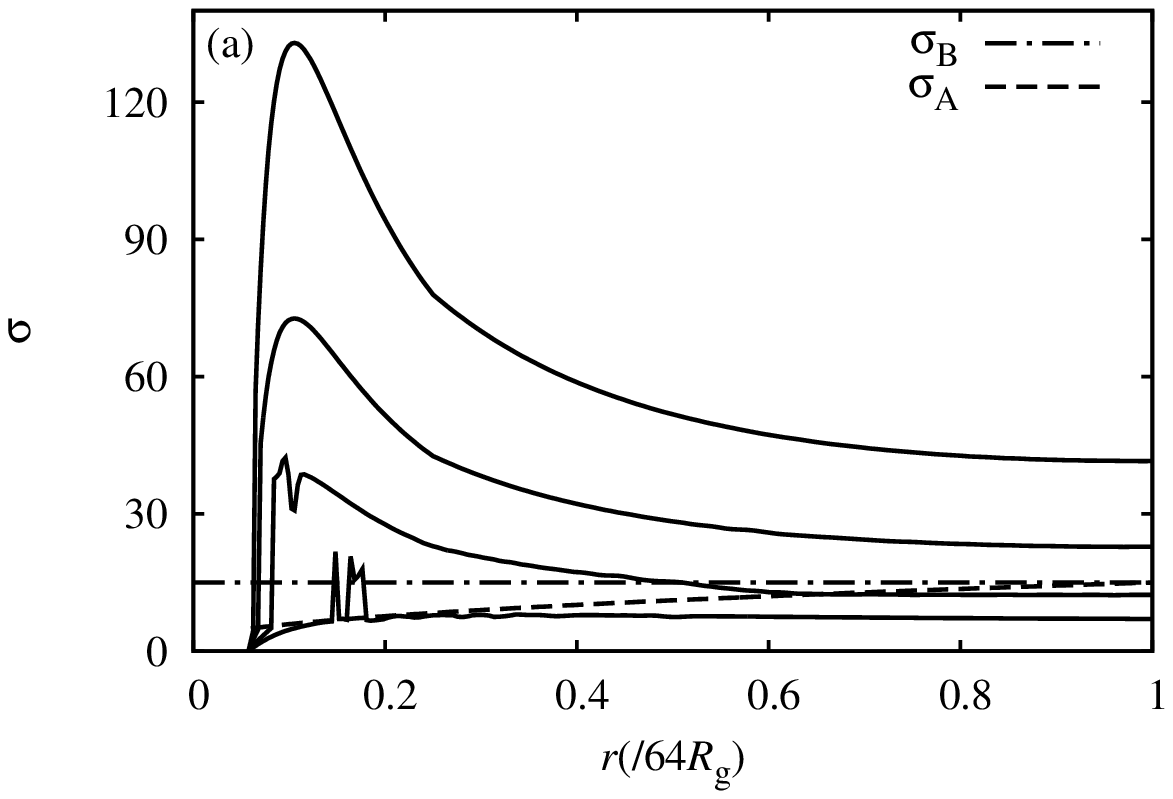}
\end{center}
\end{minipage}
\begin{minipage}{0.5\hsize}
\label{lumi-neu-rapid}
\begin{center}
\FigureFile(80mm, 50mm){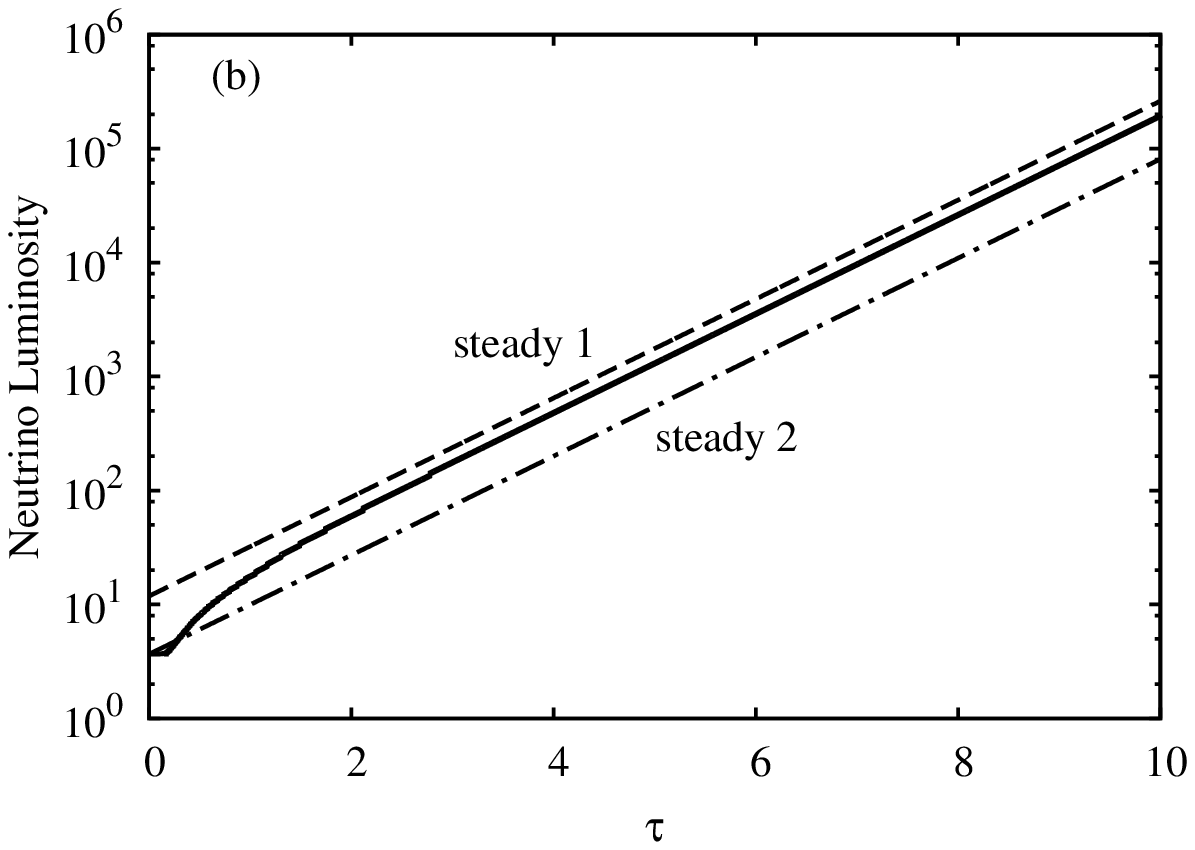}
\end{center}
%\vspace{1cm}
\end{minipage}
\caption{(a) Same as Figure 2(b) but for Model 2. The corresponding mass injection rates are $\dot{m}_{\rm inj} = \dot{m}_{0} \exp(\tau)$ with $\tau = 0.2,~0.8,~1.4$, and $2.0$. (b) Same as Figure 7(a) but for Model 2.}
\end{figure*}

\subsection{Model 3 (decay)}

\subsubsection{Surface-Density Evolution}

In the last model (Model 3), we let the mass injection rate decrease on the viscous timescale.
We compare the steady model (left) and the time-dependent model (right) in the same way as we did in Model 1 [see Figure 9].
The adopted mass injection rates are $\dot{m}_{0} \exp(-\tau)$ with $\tau=0.0,~0.8,~1.4$, and $2.0$ [see equation (\ref{mdot-inj-3})].
We clearly see in both these panels that the high-$\sigma$ region producing intense neutrino emission gradually shrinks. In addition, we notice a tendency that the high-$\sigma$ regions are systematically narrower in the steady model than in the time-dependent model. This is because it takes about the viscous timescale until the information of reduced $\dot{m}_{\rm inj}$ is transported to each portion of the disk so that it can lose the surface density to reach point B in the right panel of Figure 1.\\

\begin{figure*}
\begin{minipage}{0.5\hsize}
\label{steady-sigma-2}
%\vspace{-1cm}
\begin{center}
\FigureFile(80mm, 50mm){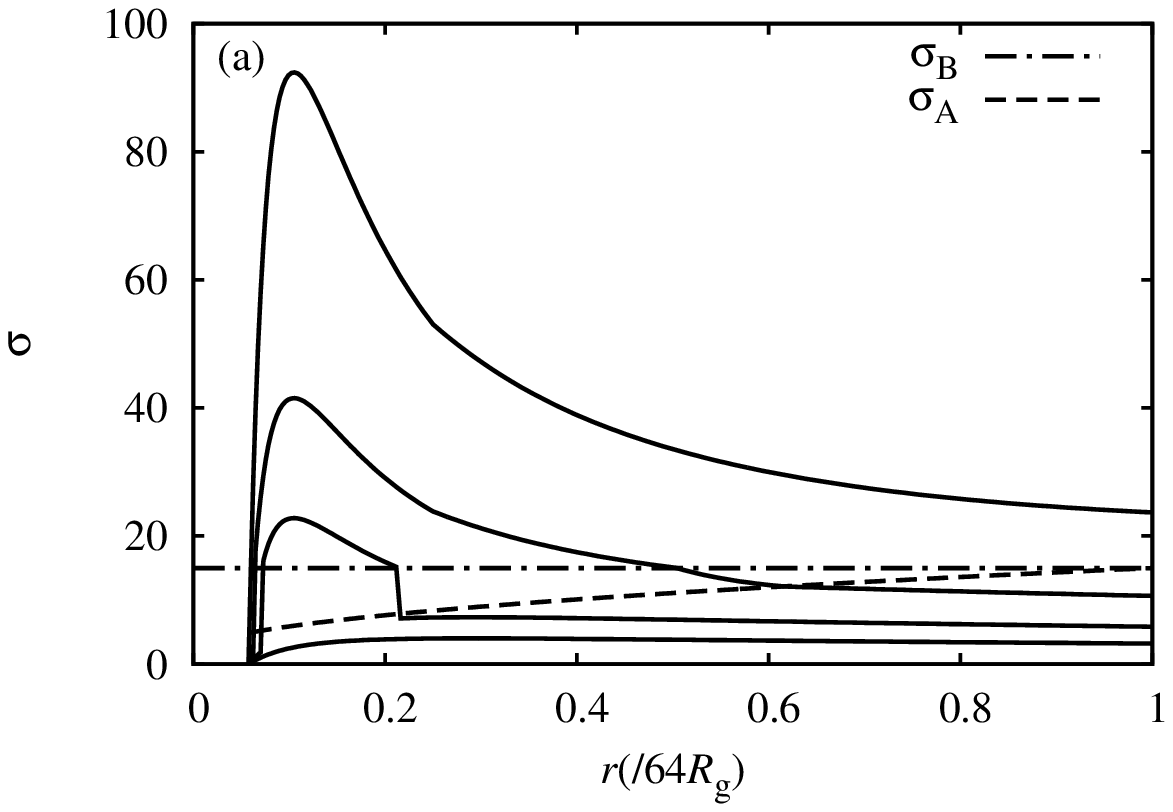}
\end{center}
\end{minipage}
\begin{minipage}{0.5\hsize}
\label{sigma-hikaku-2}
\begin{center}
\FigureFile(80mm, 50mm){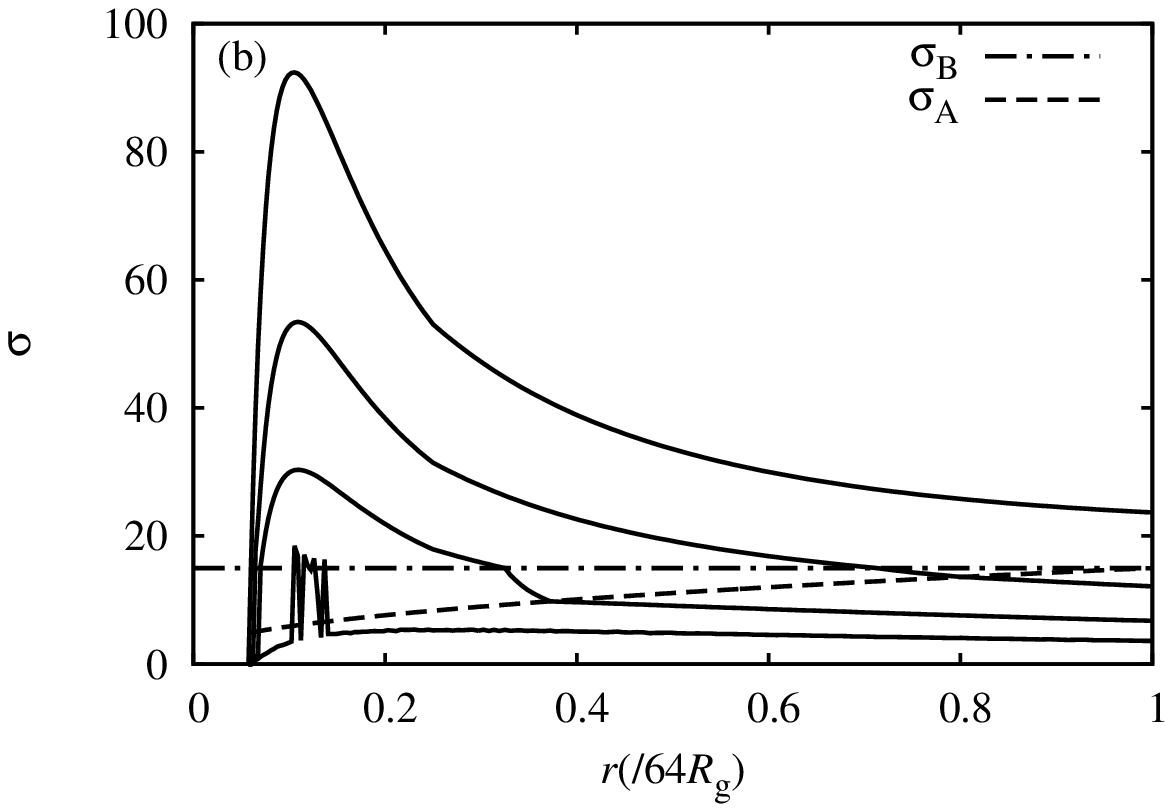}
\end{center}
%\vspace{1cm}
\end{minipage}
\caption{Same as Figure 2 but for Model 3 (decay). We compare the results of (a) the steady model and (b) the time-dependent model at the elapsed times of $\tau=0.0,~0.8,~1.4$, and $2.0$.}
\end{figure*}

\subsubsection{Mechanism of Transitions Induced by Non-steady Effects}

\begin{figure}
\begin{center}
%\vspace{-1cm}
\FigureFile(80mm, 50mm){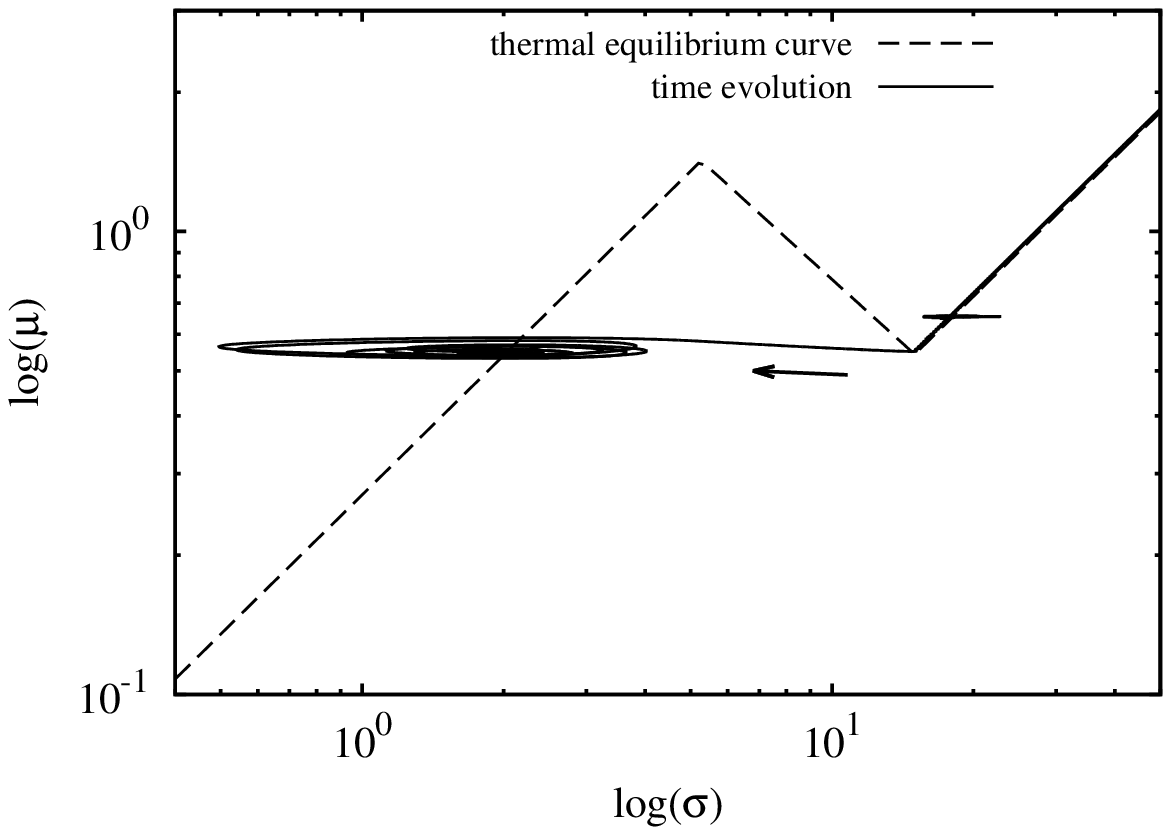}
%\vspace{1cm}
\end{center}
\caption{The actual behavior on the $\sigma$-$\mu$ plane at the radius $x=0.27$ in Model 3: The solid line represents the time evolutionary path, while the dashed line represents the thermal equilibrium curve. The arrow indicates the direction of time evolution.}
\label{nusigma54}
\end{figure}

\begin{figure*}
\begin{center}
\FigureFile(120mm, 50mm){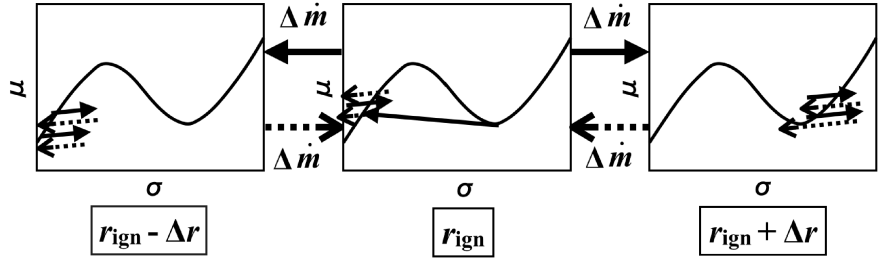}
\end{center}
\caption{Same as Figure 5(c) but for Model 3 (decay).}
\label{mangaD}
\end{figure*}

Figure \ref{nusigma54} shows the simulation results of Model 3 on the $\sigma$-$\mu$ plane, while Figure \ref{mangaD} is a schematic picture explaining the time evolution of each portion of the disk on same plane.
When point B is reached at some radius (referred to as $r_{\rm ign}$), a downward transition is triggered there, thereby the value of $\mu$ sightly increasing. This then causes non-steady mass flow ($\Delta \dot{m}$) from $r_{\rm ign}$ to the neighboring regions so that $\sigma$ should decrease at $r_{\rm ign}$.
Even after the downward transition is completed, the $\mu$ and $\sigma$ are oscillating around their equilibrium values more remarkably than in Model 1.
We explain why these violent oscillations occur in Sec. 3.4.3.
Such fluctuations in $\mu$ induce mass accretion flow from or into the neighboring regions and occasionally trigger transitions to the lower branch in the surrounding region.
It is also important to note that such propagation of the instability is restricted in a narrow zone. In other words, it cannot produce coherent oscillation.\\

\subsubsection{Wave Propagation}

\begin{figure}
%\vspace{-1cm}
\begin{center}
\FigureFile(80mm, 50mm){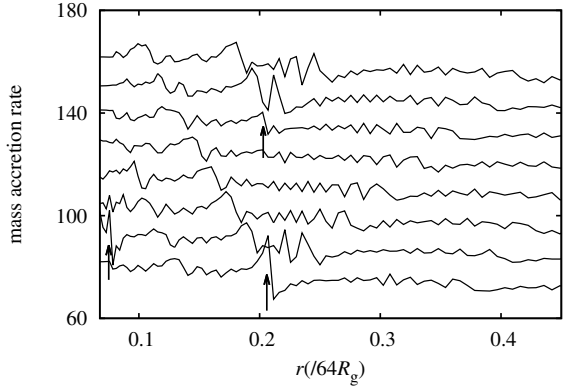}
\end{center}
%\vspace{1cm}
\caption{Same as Figure 6(a) but for Model 3. The elapsed times are 1.711, 1.714, 1.717, $\cdots$, and 1.732. Note that the curve is offset by 0, 12, 24, $\cdots$, 84 for visible clarity.}
\label{mdot-wave-3}
\end{figure}

Next, we show in Figure \ref{mdot-wave-3} the radial distribution of the mass accretion rate after the downward transition.
It is clear in this figure that the $\dot{m}$ distribution is never flat, meaning that each part of the disk at any times suffers fluctuations.
Especially, $\dot{m}$ rapidly goes up and down with a large amplitude when a transition takes place. The influence propagates in a form of $\dot{m}$ variation wave from $r_{\rm ign}$ to both in the inner and outer directions.
The situations are similar to those in Model 1 but are more pronounced in Model 3.
Fluctuations in $\dot{m}$ drive variations of $\sigma$ [see equation (\ref{basic1'})].

Here, we think the violent fluctuations mentioned in Sec. 3.4.2 over again [see Figure \ref{mangaD}]. 
The fluctuations are the results of the back-reactions of the non-steady mass flow as mentioned in Model 1.
The values of $\sigma$ and $\mu$ change in the neighboring zones after their receiving non-steady mass outflow. The changes in $\mu$ will then modify the mass flow pattern in a next moment.
The modified mass flow pattern in the next moment further changes $\sigma$ and $\mu$ values, which sometimes cause a negative feedback.
For this reason, the $\sigma$ and $\mu$ values violently oscillates around the equilibrium curve even after the thermal timescale.\\

\subsubsection{Neutrino Luminosity and Mass Accretion Rate}

We show in Figure 13(a) the neutrino luminosity in Model 3. Comparing the changes of the time-dependent luminosity with the changes of the luminosities in the steady models, we can see that the luminosity rapidly decreases when downward transitions from the upper regime (NDAF) to the lower (ADAF) branch take place.

We plot the accretion rate at the inner edge of the disk for Model 3 in Figure 13(b). We again see the significant variability in the mass accretion rate, as we saw in Model 1.

\begin{figure*}
\begin{minipage}{0.5\hsize}
\label{case3-lumi-neu}
%\vspace{-1cm}
\begin{center}
\FigureFile(80mm, 50mm){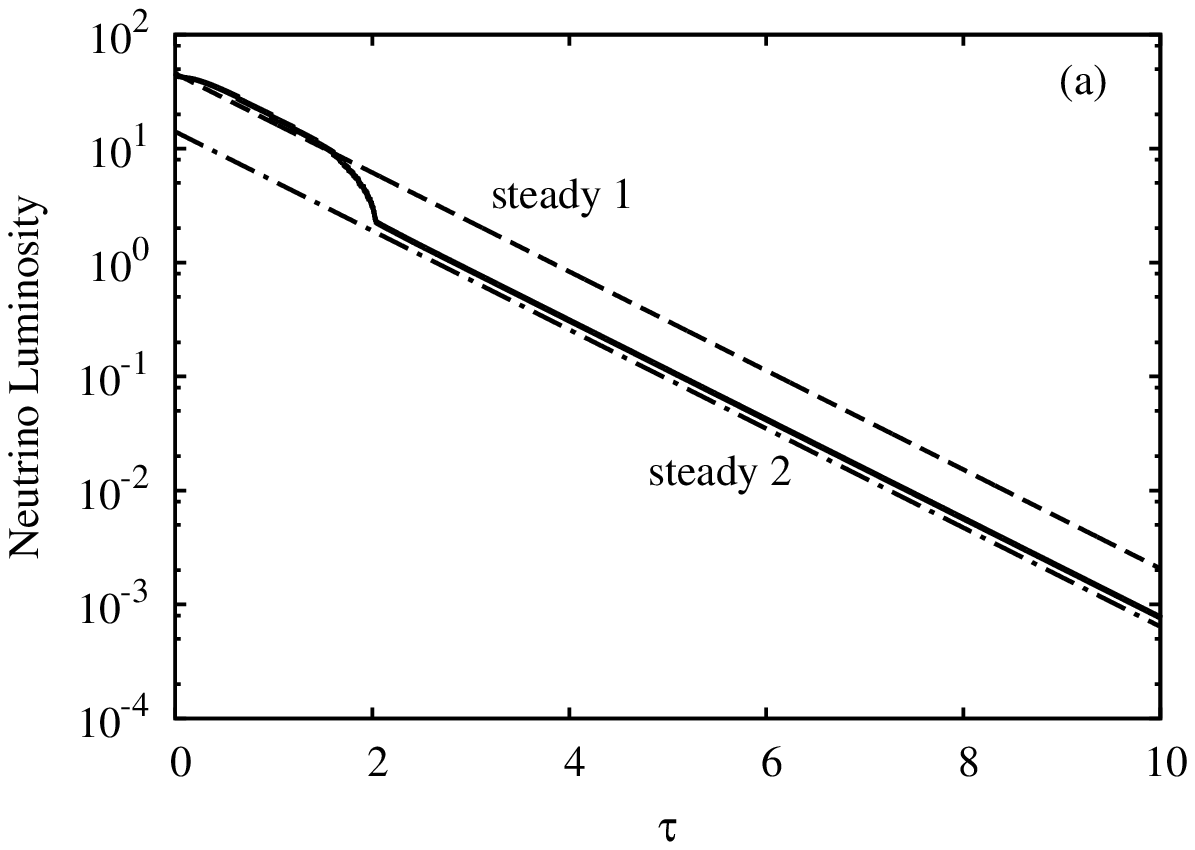}
\end{center}
\end{minipage}
\begin{minipage}{0.5\hsize}
\label{mdot-acc-3}
\begin{center}
\FigureFile(80mm, 50mm){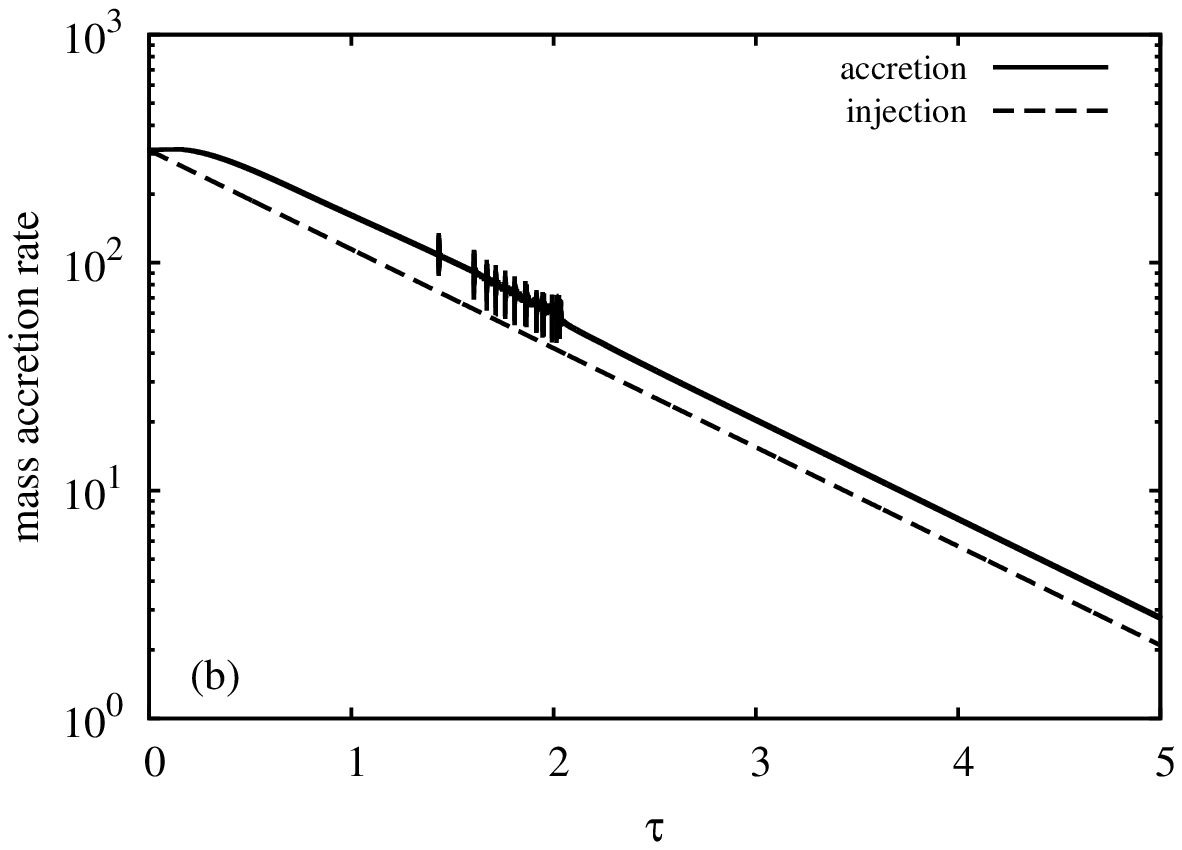}
\end{center}
%\vspace{1cm}
\end{minipage}
\caption{Same as Figure 7(a) and (b) but for Model 3.}
\end{figure*}

\section{Brief Summary and Discussion}

\subsection{Brief Summary}

The thermal structure of the hyperaccretion flows is characterized by an N-shaped thermal equilibrium curve: the middle branch, which is secularly unstable, is sandwiched by the lower ADAF branch and the upper NDAF branch.
In this study, we examine the time evolution of a disk, which has N-shaped thermal equilibrium curves, by means of simple numerical simulations.
We solve time-dependent, height-averaged basic equations for viscous disks receiving with variable mass injection to see how the transition between the lower and upper branches occurs, how it propagates over a disk plane, and what differences arise from the cases without N-shaped equilibrium curves.
We studied three models, which have different time dependence of the mass injection rate: Model 1 (slow rise), Model 2 (rise), and Model 3 (decay).
What we find can be summarized as follows: 

\begin{enumerate}
\item 
When the mass injection rate increases (or decreases) with time, the region initially on the lower (upper) branch successively undergoes upward (downward) transitions.
The transition associated with rapid changes in the surface density and propagates over the disk, giving rise to an expansion (shrinkage) of the neutrino-emission region.
\item 
Non-steady mass flow ($\Delta \dot{m}$), the mass flow deviated from the spatially averaged flow, arises in a narrow region around the ignition radius $r_{\rm ign}$ during a transition.
This is because the middle branch is secularly unstable; that is, the larger (or the smaller) the surface density of the unstable region is, the more (less) mass enters the ignition region than its neighboring regions.
Resultant $\dot{m}$ variation wave propagates over the whole disk.
\item 
Such non-steady effects can occasionally cause transitions even in the region at which the transition would not occur, if the disk were in a steady state. The more slowly the mass injection rate changes, the more remarkable the non-steady effects become.
The effects are, however, not strong enough to produce coherent transition over the entire disk plane.
\item
The transition between the lower and upper branches yields abrupt changes in the neutrino luminosity. In fact, the neutrino luminosity calculated based on the time-dependent results changes much more rapidly than that in the case without N-shaped equilibrium curves. Non-steady effects can be seen as fluctuations of the neutrino luminosity and in mass accretion rates.
\end{enumerate}

\subsection{Effects of the modified N-shape on disk evolution}

We have argued that the instability is a source of producing the non-steady effects.
If so, the properties of the disk evolution may critically depend on the precise shape of the N-shaped thermal equilibrium curve.
In this subsection, we modify the thermal equilibrium curve in such a way to enhance the instability to see if it is possible for the instability to propagate more widely on the disk plane.

\begin{figure}
%\vspace{-1cm}
\begin{center}
\FigureFile(80mm, 50mm){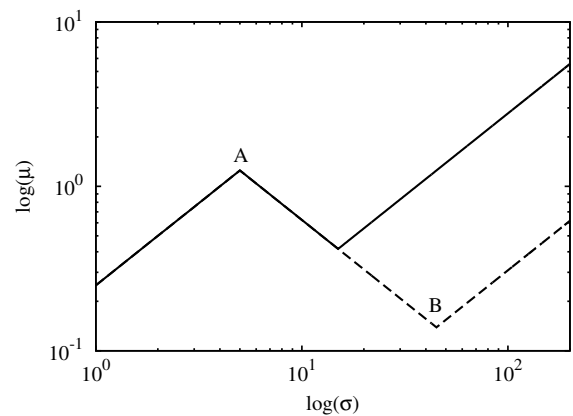}
\end{center}
%\vspace{1cm}
\caption{Modification of the N-shaped equilibrium curve at $r=4R_{\rm g}$: The solid line is the original curve while the dashed line is the modified one.}
\label{n-curve-hikaku}
\end{figure}

\citet{kawanaka+13b} adopted the $\alpha$ viscosity prescription in calculating the thermal equilibrium curve of a hyperaccretion shown in the left panel of Figure 1. Although they set $\alpha$ to be constant (= 0.1), there are no strong reasons to believe so.
As a thought experiment we set $\alpha = 0.01$ on the upper branch, keeping the location of the lower branch and the slope of the middle branch unchanged.
The value of $\sigma_{\rm B}$ is now 45.0, instead of 15.0 (see Figure \ref{n-curve-hikaku}).
By this change, the middle unstable branch in the equilibrium curve is more elongated than in the original model. This model is referred to as the modified N-shape model.

\begin{figure}
%\vspace{-1cm}
\begin{center}
\FigureFile(80mm, 50mm){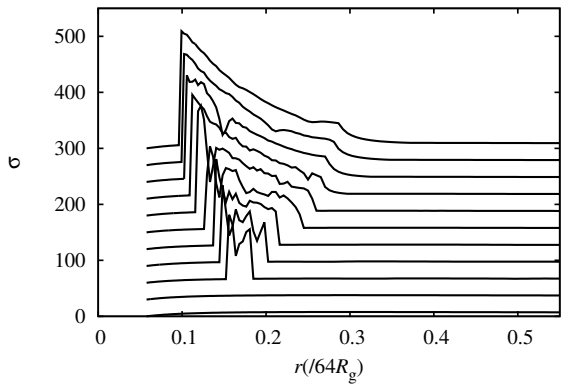}
\end{center}
%\vspace{1cm}
\caption{Same as Figure 3(b) but for the modified N-shape model.
The elapsed times are $\tau=0.0-3.0$ with a constant interval of $\delta \tau=0.3$. Each line is offset vertically by the value of $100 \times \tau$ for visible clarify.}
\label{sigma1-d}
\end{figure}

\begin{figure*}
\begin{minipage}{0.5\hsize}
\label{lneu-case1-d}
%\vspace{-1cm}
\begin{center}
\FigureFile(80mm, 50mm){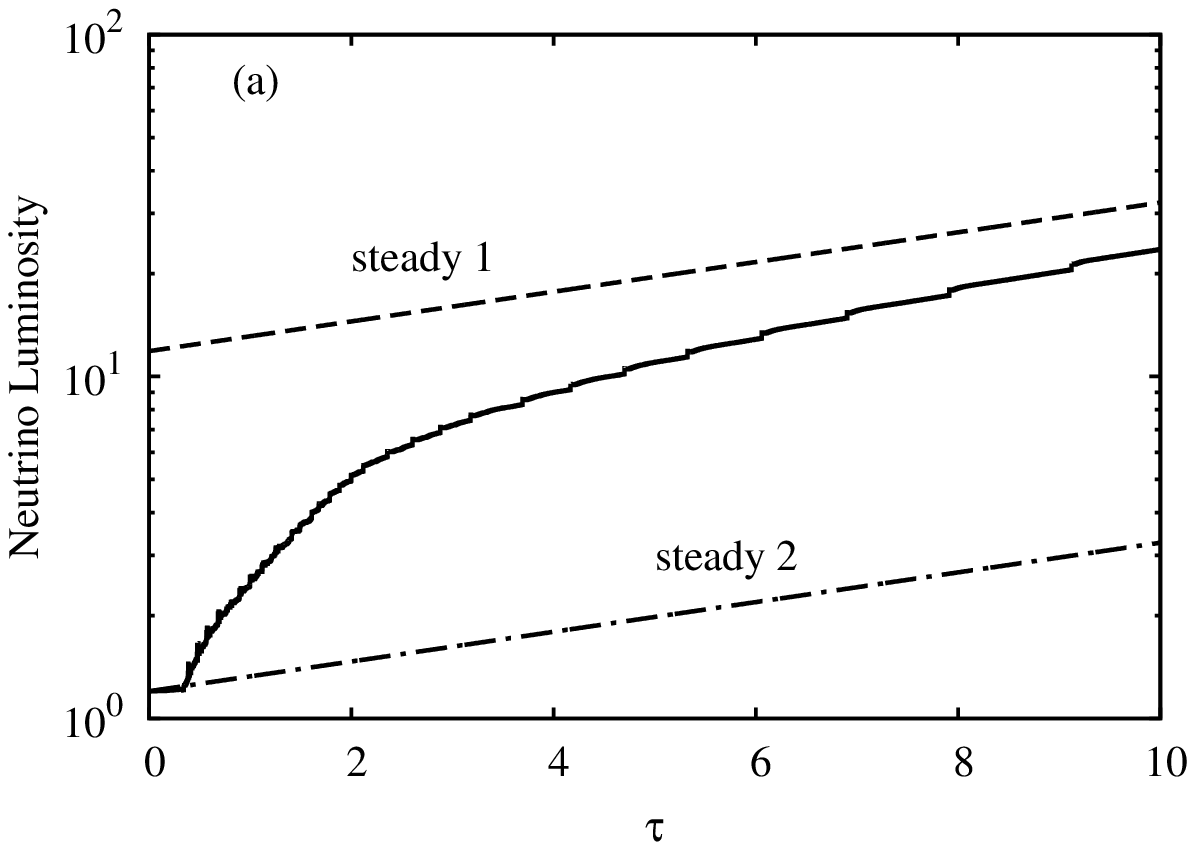}
\end{center}
\end{minipage}
\begin{minipage}{0.5\hsize}
\label{acc-1d}
\begin{center}
\FigureFile(80mm, 50mm){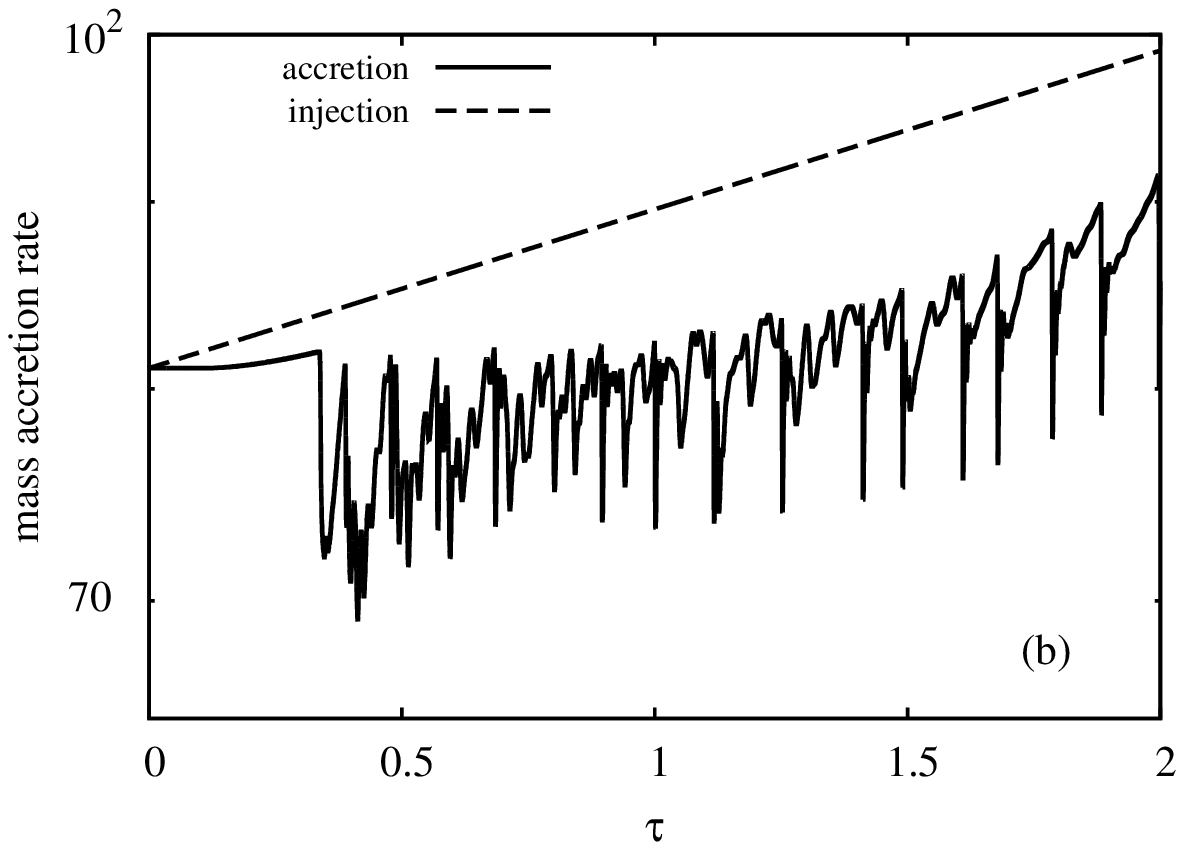}
\end{center}
%\vspace{1cm}
\end{minipage}
\caption{Same as Figure 7(a) and (b) but for the modified N-shape model.}
\end{figure*}

As to the mass injection rate, we only calculate the case of slow rise [i.e., Model 1 in section 3; see also equation (\ref{mdot-inj-1})].
First, we show the global evolution of the surface-density distribution in Figure 15.
We can see that the region expands in a similar way to that in Figure 3(b). Note also that the high-$\sigma$ region with efficient neutrino emission values in the upper branch are higher.
Next, we calculate the luminosity and show it in Figure 16(a). Comparing it with Figure 7(a), we understand that the modification of the N-shape does yield more rapid increase in neutrino luminosity, but the speed of the upward transition wave is not so much affected.
We also calculate the mass accretion rate at radius $r_{\rm in}$ in Figure 16(b). Comparing it with Figure 7(b), we find that the accretion rate varies more or less continuously.
These variations look similar to the observed ones in gamma-ray bursts (see the next subsection).

\subsection{Implications on the GRB Emission Feature}

Here we briefly discuss the observable features of the unstable hyperaccretion flows.  The short-term variability of a hyperaccretion flow would be the origin of the highly variable prompt emissions of GRBs because the variable energy deposition from an accretion flow would produce a spatially inhomogeneous outflow, and it would result in producing internal shocks, which give rise to the highly variable emission from shock-accelerated electrons.

According to our calculations, the neutrino luminosity shows abrupt changes with time during the transition of a hyperaccretion flow.  In Model 1 with slowly increasing mass injection, especially, the step-function-like changes in the neutrino luminosity occur repeatedly.  If we adopt the neutrino pair annihilation scenario as the energy deposition process from a hyperaccretion flow, we can expect the intermittent energy ejection from an accretion flow, which can result in the highly variable emission of GRBs.

As for Model 3, where the injection rate decreases with time, the mass accretion rate at the inner edge shows strong variability.  This may be related to the short-term variation of the Blandford-Znajek jet luminosity since it is proportional to the magnetic energy density on the event horizon and is, hence, likely to be proportional to the mass accretion rate (Kawanaka et al. 2013a).

We wish to note that the behavior of the secular instability is not always controlled by the variation timescale of the mass injection rate, but that rapid variations can occur on a much shorter timescale, less than $0.1~t_0 \sim 10^{-2}$ s [see figure 7(b) and equation (\ref{t0})]. 
We here need to distinguish two timescales: the timescale of variable mass injection rate (which is roughly the viscous timescale at large radii) and that of the fluctuations in the $\sigma$ and $\mu$ values around the ignition point (which is much shorter than the former). 
The variability timescale observed in GRBs is the order of $\lesssim 10^{-2}~{\rm s}$ (e.g., \cite{ackermann+10, troja+15}).
The observed rapid variability of the GRBs should be related to the latter process, and so we conclude that the variability timescale in our model can match the observations of GRBs.

\subsection{Comparison with the Dwarf-Nova Type Instability}

It is well known that the outbursts of dwarf-novae, a subclass of cataclysmic variables, and probably X-ray nova eruptions are triggered by a disk instability caused by the partial ionization of hydrogen in the disk (see, e.g., Kato et al. 2008, chapter 5 and references therein).
This is the so-called dwarf-nova type instability (or hydrogen ionization instability) and is characterized by the S-shaped equilibrium curve.
It can be easily shown that the middle branch is thermally and secularly unstable, while the other two branches are stable. Under the circumstances relaxation oscillations between the bi-modal stable branches spontaneously occur, even if the mass injection rate does not vary in time.
This is the most successful time-dependent theory of the viscous accretion disk.

There are similarities and differences between the dwarf-nova type instability and the present case of hyperaccretion disk instability with the N-shaped equilibrium curve. The most prominent common feature is bi-modal nature of the equilibrium curve.
There are three states for a given $\dot{m}_{\rm inj}$ and the middle state being secularly unstable in the present case, while dwarf-nova type disks undergo bi-modal transitions, although only one state is found for a given $\dot{m}_{\rm inj}$. Another common feature is global propagation of bi-modal transitions from the ignition point to its neighboring sites.

A big distinction resides in the fact that the solution is only secularly unstable in the present case. This is a crucial point when discussing the global nature of the instability in the present case, since the thermal timescale is much shorter than the viscous timescale, on which a secular instability grows.
We in fact find non-steady features not so much pronounced, judging from the fact that the transition is restricted in a narrow region and light curves exhibit small-amplitude fluctuations. This contrasts the case in the dwarf-nova type disks which undergoes coherent oscillation and produce large-amplitude luminosity variations.
To summarize, the big distinction arises from the fact that the $\mu$ value can change with a large amplitude, thereby producing significant non-steady effects in a thermally unstable disk, while it cannot in a thermally stable disk.

\subsection{Study in the Future}

It might be important in this respect that we have employed one-dimensional formulation for the basic equations in the present study, assuming that the vertical structure of the disk can immediately respond to changes in $\sigma$.
This is not precisely correct, since it takes about a sound crossing time over the disk height ($\sim H/c_{\rm s}$) for the entire vertical structure to adjust to any changes in $\sigma$. In future work, we thus need two-dimensional simulations to see vertical propagation of the instability.

Further, discontinuous transitions in the neutrino light curve, which seems to depend on a finite mesh size, can be altered by considering the two-dimensional effects.
This is because changes in the vertical structure take longer time than the propagation time of the instability in the radial direction over a distance of one mesh size.
We naively expect that the region with the width of $\sim H$ may simultaneously undergo the branch transition, if we properly consider the two-dimensional effects.

In order to solve the time evolution of a hyperaccretion disk strictly, we have to solve equation (\ref{thermal.eq}) without modeling it. In other words, we have to evaluate advection cooling rate ($Q_{\rm adv}^{-}$) as the function of the radial velocity ($v_{\rm r}$) and the $r$ differentiation of the entropy ($ds / dr$) and neutrino cooling rate ($Q_{\nu}^{-}$) as the function of the temperature ($T$).

\vspace{8mm}

We are grateful to the anonymous referee for insightful comments.
This work is supported in part by Grants-in-Aid of the Ministry of Education, Culture, Sports, Science and Technology (MEXT) (26400229 SM).

\vspace{8mm}

\appendix

\section{Numerical Procedures}

In our calculations, we use the implicit method for time integration of
equations (\ref{basic1'}) and (\ref{basic2'}). At each time step, we solve these equations, as well as the energy equation in non-dimensional forms [see equation (\ref{thermal.eq})]. The radial coordinate ($x$= 0.0 -- 1.0) is divided into 200 mesh points (with a constant mesh spacing of $\Delta x = 0.005$); $x_i=(\Delta x)i$ with $i=1, 2, \cdots, 200$.
We distinguish cell variables (such as $\sigma$ and $\mu$) and interface variables (such as $\dot m$). The former variables are defined at the same place of the radial mesh points; e.g., $\sigma_i$ is the surface density at $x_i$ ($i=1, 2, ... 200$), while the latter variables are defined between the mesh points; e.g., $\dot m$ is the mass flow rate from the cell at $x_{i+1}$ to the cell at $x_{i}$.

Second, we set $\Delta\tau$ to be constant ($=10^{-6}$) being independent of $x$. In the present study we assign $\Delta\tau = 10^{-6}$ so as to satisfy the two conditions, $\Delta\sigma / \sigma < 0.03$ and $\Delta\mu / \mu < 0.03$ at every radius.
When either of these conditions is unsatisfied at some mesh point(s), we re-calculate by using smaller $\Delta\tau$ so that conditions should be met.
One exception is Model 3, in which we omit calculations in the innermost region (at $x \leq 0.26$) to avoid numerical difficulties.

\section{Accuracy of Numerical Calculation}

As a test of the numerical code, we check the mass conservation, we compute $M_{\rm disk}$, the disk mass in the following two methods and the difference between them. We use trapezoidal rule for this integral calculus. In method A we calculate

\begin{equation}
M_{\rm disk}^{\rm A} = 2\pi\int_{r_{\rm in}}^{r_{\rm out}} r \sigma dr = 4\pi\int_{x_{\rm in}}^{x_{\rm out}} x^{3} \sigma dx.
\label{method-A}
\end{equation}
In method B we calculate
\begin{equation}
M_{\rm disk}^{\rm B} = \int_{0}^{\tau_{\rm f}} (\dot{m}_{\rm inj} - \dot{m}_{\rm acc}) d\tau.
\label{method-B}
\end{equation}

We have confirmed that the total mass is conserved within the accuracy of less than 0.2$\%$ in Model 1, less than 0.8$\%$ in Model 2, and less than 3.6$\%$ in Model 3 for time integration of $\tau = 0 - 2.0$ ($\tau_{\rm f}=2.0$).

\end{document}